\documentclass[10pt]{report}
\usepackage[english]{babel}
\usepackage[utf8x]{inputenc}
\usepackage{amsmath}
\usepackage{graphicx}
\usepackage[colorinlistoftodos]{todonotes}
\usepackage[toc,page]{appendix}
\usepackage[a4paper, total={6in, 9in}]{geometry}
\usepackage{amssymb}
\usepackage{pifont}
\newcommand{\cmark}{\ding{51}}%
\newcommand{\xmark}{\ding{55}}%
\usepackage{afterpage}
\usepackage{hyperref}
\usepackage{titlesec, lipsum}
\usepackage{titling}
\usepackage[nottoc,numbib]{tocbibind}
\usepackage{caption}
\usepackage{subfig}
\usepackage{listings}
\renewcommand\lstlistlistingname{List of Listings}

\hypersetup{
    colorlinks,
    citecolor=blue,
    filecolor=blue,
    linkcolor=blue,
    urlcolor=blue
}

\begin{document}

\begin{titlepage}

\newcommand{\HRule}{\rule{\linewidth}{0.5mm}} 
\setlength{\topmargin}{0in}
\center 

 \begin{minipage}{0.4\textwidth}
\begin{flushleft} \large
\vspace*{-2cm}
\hspace*{-1cm}
\includegraphics[scale=1]{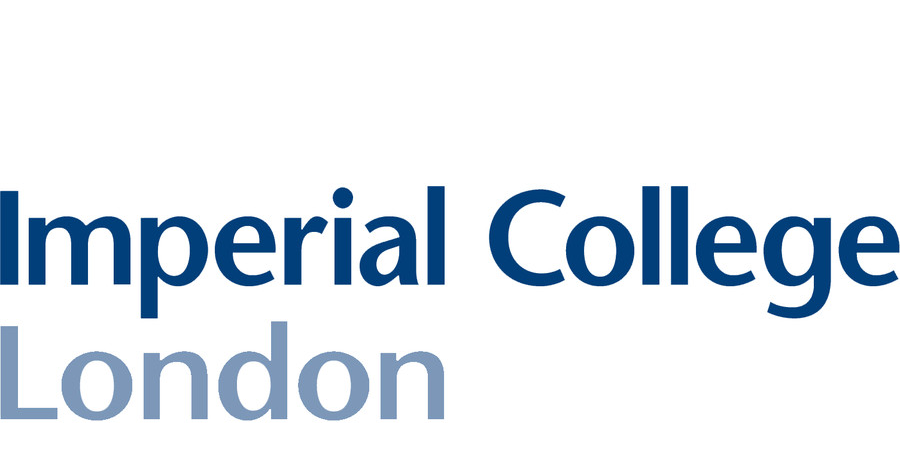}\\
\end{flushleft}
\end{minipage}
~
\begin{minipage}{0.5\textwidth}
\begin{flushleft} \large
\vspace*{-0.5cm}
\hspace*{5cm}
\includegraphics[width=4cm]{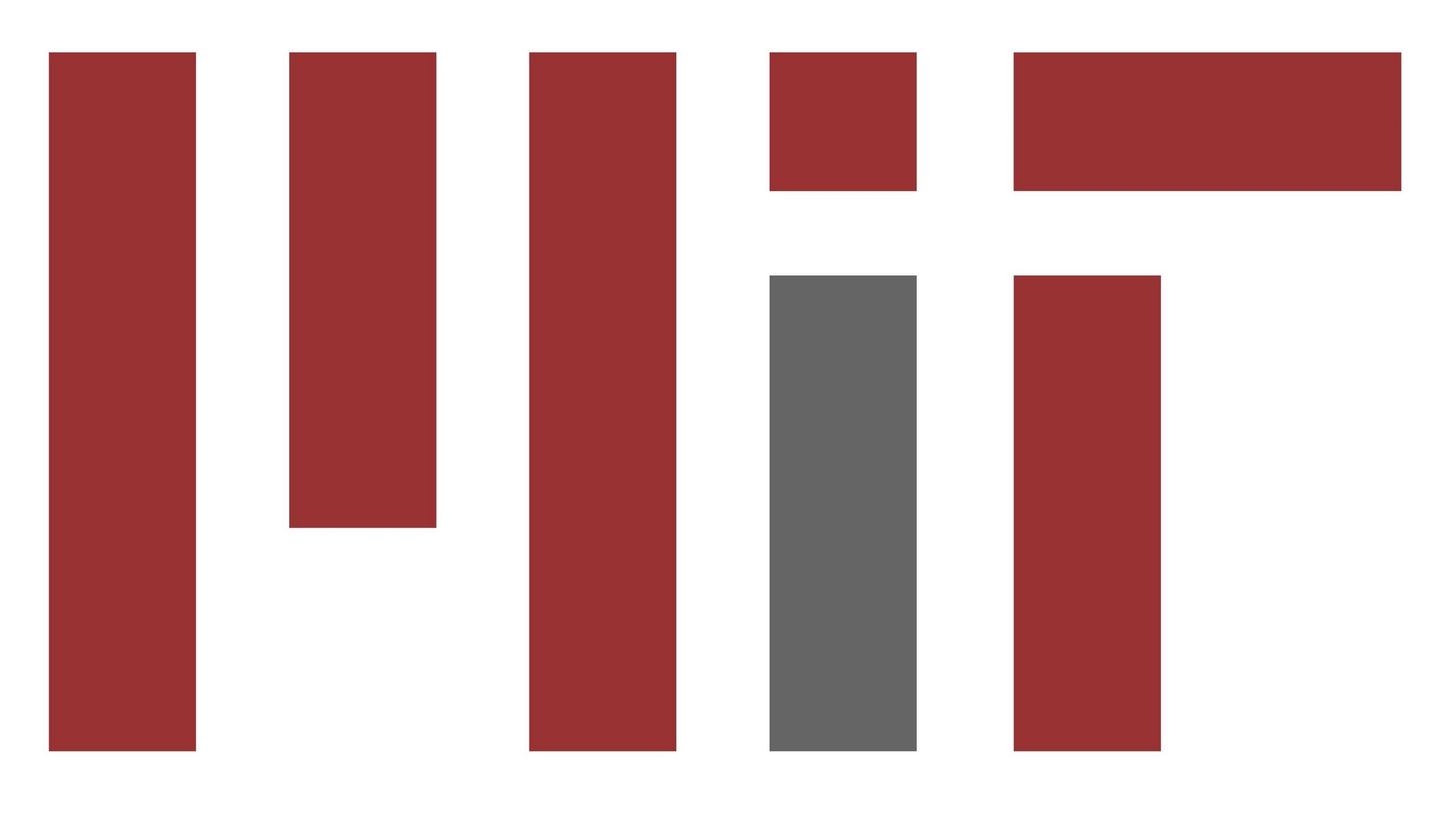}\\
\end{flushleft}
\end{minipage}\\[1cm]

\textsc{}\\[2cm]
\textsc{\Large MEng Individual Project}\\[1cm] 

\textsc{\large Imperial College of Science, Technology and
Medicine}\\[0.5cm]
\textsc{\large Massachusetts Institute of Technology}\\[1cm]
\textsc{\Large Department of Computing}\\[0.5cm] 


\HRule \\[0.4cm]
{ \huge \bfseries Modelling Concurrency Bugs Using Machine Learning}\\[0.4cm] 
\HRule \\[1cm]
 

\begin{minipage}{0.4\textwidth}
\begin{flushleft} \large
\emph{Author:}\\
Teodor-Rares \textsc{Begu} \\ 
\end{flushleft}
\end{minipage}
~
\begin{minipage}{0.5\textwidth}
\begin{flushright} \large
\emph{Mentor:} \\
Shashank \textsc{Srikant} \\[0.5cm] 
\emph{Faculty Supervisor:} \\
Dr. Una-May \textsc{O'Reilly} \\[0.8cm]

\emph{Second markers:} \\
Dr. Sergio \textsc{Maffeis} \\[0.0cm]
Dr. Anandha \textsc{Gopalan} \\[0.2cm]
\end{flushright}
\end{minipage}\\[1cm]

{\large May 29, 2020}\\[0.5cm]

\vfill 
\afterpage{\null\newpage}

\end{titlepage}

\setcounter{page}{3}

\begin{abstract}
Artificial Intelligence has started to gain a lot of traction in the recent years, with machine learning notably starting to see more applications across a varied range of fields, in areas as diverse as healthcare \cite{healthcare}, finance \cite{finance} and self-driving cars \cite{selfdriving}.

 One specific machine learning application that is of interest to us is that of software safety and security, especially in the context of parallel programs. The issue of being able to detect concurrency bugs automatically has intrigued programmers for a long time \cite{savage1997eraser}, \cite{chamillard1996empirical}, as the added layer of complexity makes concurrent programs more prone to failure. The development of such automatic detection tools provides considerable benefits to programmers in terms of saving time while debugging, as well as reducing the number of unexpected bugs. 
 
 We believe machine learning may help achieve this goal by providing additional advantages over current approaches, in terms of both overall tool accuracy as well as programming language flexibility. However, due to the presence of numerous challenges specific to the machine learning approach (correctly labelling a sufficiently large dataset, finding the best model types/architectures and so forth), we have to approach each issue of developing such a tool separately. 
 
 Therefore, the focus of this project is on comparing both common and recent \cite{zaheer2017deep} machine learning approaches. We abstract away the complexity of procuring a labelled dataset of concurrent programs under the form of a synthetic dataset that we define and generate with the scope of simulating real-life (concurrent) programs. We formulate hypotheses about fundamental limits of various machine learning model types which we then validate by running extensive tests on our synthetic dataset. We hope that our findings provide more insight in the advantages and disadvantages of various model types when modelling (concurrent) programs using machine learning, as well as any other related field (e.g. Natural Language Processing).
\end{abstract}

\newpage

\setcounter{page}{4}

\renewcommand{\abstractname}{Acknowledgements}
\begin{abstract}
First and foremost, I would like to thank my parents for supporting me throughout my university years and encouraging me to pursue my passions no matter what they might be.

I would like to offer special thanks to my mentor Shashank Srikant and faculty supervisor Dr Una-May O'Reilly for offering me this fantastic research opportunity and guiding me throughout this project.

I would also like to thank Dr Sergio Maffeis for coordinating this exchange year between Imperial and MIT without which I would not have had the opportunity to work on this project, let alone have the fascinating experience of studying at MIT.

\end{abstract}

\newpage

\setcounter{page}{5}

\tableofcontents
\newpage

\listoffigures
\newpage

\listoftables

\newpage

\addcontentsline{toc}{chapter}{\lstlistlistingname}{\lstlistoflistings}
\newpage

\chapter{Introduction}

\begin{figure}[!htbp]
    \centering
    \includegraphics[scale=0.45]{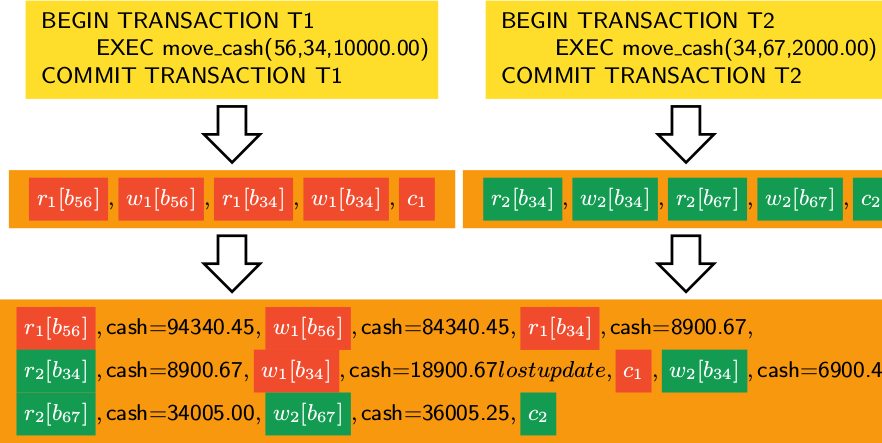}
    \caption{Taken from CO572 lecture notes \cite{lostupdate}, this example showcases a specific concurrency bug (lost update). Because of a specific interleaving of the two threads, Thread 1's write to account 34 is forgotten and thus account 34's balance is down 10000 more than it should be.}
    \label{fig:lost_update}
\end{figure}

Concurrency bugs in all their forms are especially hard to find as any amount of parallelism in a program will cause the sequences of executed instructions to become non-deterministic. Since concurrent threads can interleave in any manner at runtime (one such buggy interleaving is shown below \ref{fig:lost_update}), the number of possible thread interleavings grows exponentially. Therefore, this makes it hard for analyser tools to predict bugs in code, as they cannot search all interleaving possibilities exhaustively. 

Current approaches are divided under static analysers \cite{wang2005static, vakilian2011keshmesh} and dynamic analysers \cite{flanagan2004atomizer, savage1997eraser}, with most approaches (regardless of being static or dynamic) being rule-based. In fact, some of these approaches \cite{wang2005static, flanagan2004atomizer} expand on previous work relating to the formal definition of a type system for concurrent programs \cite{flanagan2003type, boyapati2001parameterized, boyapati2002type}.

There has also been some recent work \cite{tehrani2019deeprace} on applying machine learning to this particular problem. However, we believe that machine learning models are capable of picking up more complex patterns than the ones shown there. Additionally, there have been other applications of AI or statistical models to software, with the most notable one being the prediction of program properties such as variable names or types \cite{raychev2015predicting, alon2018general, allamanis2014learning}. Another more recent paper \cite{allamanis2018survey} has a thorough look at all the related work on the intersection between AI and programming languages. \\

\section{Motivation}
One common pitfall rule-based tools have is the lack of flexibility. Developing such a tool requires technical expertise for a specific programming language, and once that tool was made, the effort cannot quickly transfer to another programming language. This proves especially troublesome for newer languages like Solidity \cite{solidity} that cannot have a ready-made concurrency detection tool. \\ 

We believe that this issue could be solved through following a machine learning approach, as models could learn concurrency patterns that are common to all languages. However, this would require the models be trained on a common intermediate representation of the languages (for e.g. the LLVM intermediate representation \cite{lattner2004llvm}, which is a lower-level representation of code that sits between the source code and the assembly code). Although, in this case, each language would require a specific translation from source to the intermediate representation. Nevertheless, we believe this to be more feasible than developing a new concurrency detection tool from scratch. Additionally, each new language would benefit from concurrency patterns previously discovered in datasets of other languages. \\

This brings us to another advantage the machine learning approach might have, which is that of accuracy. Ideally, we envision that given a large enough dataset combining multiple programming languages, machine learning models would be able to outperform rule-based tools as they could detect buggy patterns that are too unexpected or too complex for a human to write rules against (as is the case with current approaches). \\

As it is with most machine learning tasks, the data is often the most crucial aspect of the problem. However, in our context, this proves to also be the most challenging aspect. Since it takes much manual effort to correctly classify a single concurrent program as buggy or not (labelling very complex programs might not even be feasible), labelled datasets of concurrent programs are thus very hard to come by. Moreover, there is more uncertainty to the question of whether there exist machine learning models that can successfully learn patterns from code than there is to the question of whether a labelled dataset could be procured. The uncertainty of the former question combined with the laborious work of the latter question gives us the objective of this SuperUROP project. Therefore, we hope to help clarify the former question, while also providing insight into model learning capabilities for tasks with inputs similar in nature to this one. \\

In this project, we analyse different machine learning model types and architectures, formulating hypotheses about model learning capabilities and running experiments on a synthetic dataset to validate our initial reasoning. The purpose of the synthetic dataset is to emulate real-life (concurrent) programs while abstracting away the complexity of working with real programs. The benefit of using a generated synthetic dataset is thus twofold, first being the availability of training data. On the other hand, having complete control over the dataset also lets us abstract away other issues related to modelling code as input to neural networks (like variable size input, figuring out which intermediate representation to use and so forth). Thus not only can we tackle each code modelling problem one step at a time, but we can also have fine-grained control over which samples the models will be trained on and which samples they will be tested on, allowing us to stress test model limits and capabilities extensively. 

\section{Contributions}

We generally test four types of artificial neural networks, namely simple feedforward neural networks with only dense layers, convolutional neural networks (CNN's), recurrent neural networks (RNN's) (under the form of long short-term memory networks, or LSTM's) and a recent neural network architecture called DeepSets \cite{zaheer2017deep}. Our results confirm our hypotheses that only CNN's and LSTM's so far can generalise when being given only a specific subset of the total available dataset. Across all of our experiment variations, CNN and LSTM model types consistently achieve an accuracy of above 90\% (with an overall accuracy across all experiments of 98.4\% and 98.6\% respectively), with feedforward and DeepSets only achieving 78.2\% and 78.1\%  respectively. Future work along these lines would also include testing more RNN variants like gated recurrent unit networks (GRU's) \cite{chung2014empirical} or graph neural networks \cite{zhou2018graph}.\\

Our key contributions are thus summarised as follows:
\begin{itemize}
    \item \textit{Generation of a synthetic labelled dataset meant to simulate patterns found in real-life (concurrent) programs.} We define a formal toy programming language using BNF grammar, which we then use to generate simulated programs. We then define underlying rules by which a simulated program generated from this language is classified as buggy or not. Although the rules are simple for a human to understand and interpret, the models must also demonstrate this capability.
        
    \item \textit{Filters to separate the dataset into train and test: by careful separation, we can test for specific properties of the models.} Much like in all machine learning tasks, seeing only certain types of samples during training might mean the models overfit and have poor accuracy during testing. Having these filters allows us to test the models' generalising ability as well as robustness to overfitting by selectively determining which samples they can only see during training and which samples they can only see during testing.
    
    \item \textit{Reasoning about which properties are to be expected for each machine learning model type, validated empirically by evaluating the models on the synthetic dataset.} We formulate a set of desirable properties and reason about which models might exhibit them and why. We define tests for these properties by using specific filters, and we then look at the accuracy of the models on those filters.
    
\end{itemize}

\newpage
\chapter{Background}
\section{Concurrency}

Concurrency refers to the ability of a program to have multiple sections of code (threads) executing in parallel at the same time \cite{concurrencywiki}. For instance, this can be utilised to improve execution speeds of computation heavy programs as well as provide a smoother user experience in applications that also have a graphical user interface component. Concurrent programs do not have a deterministic sequential execution, unlike their non-concurrent counterparts, as it is highly unlikely that a program has the same interleaving of threads for different executions. This can be due to a plethora of factors, such as relative speeds of thread execution, thread scheduling by the operating system \cite{threadswiki}, or dependence on external resources or input. 

This non-deterministic property greatly affects the reproducibility \cite{reproducibilitywiki} of automated tests on concurrent programs, which in turn affects the reliability of concurrent programs. It might be that a concurrent program passes all automated tests, but during production, there is a different thread interleaving which causes a bug in the program. Since writing automated tests is thus not as reliable, the need for tools that can accurately detect concurrency issues is thus justified.

\subsection{Synchronization mechanisms}
Synchronisation mechanisms refer to methods through which the programmer can coordinate access of multiple threads to shared memory within the program, ensuring no more than one thread at a time can enter what is called the \textit{critical section} \cite{synchmechwiki}. Whenever synchronisation methods correctly ensure mutually exclusive access to a critical section, the section of code representing the section thus becomes atomic, which ensures non-interference from other threads \cite{flanagan2003type}. However, misuse of synchronisation mechanisms or the lack of use thereof can lead to concurrency problems as described in the next section.

There exist a plethora of synchronisation methods depending mostly on the availability in the source language, ranging from high-level synchronised methods in Java \cite{javasynch} to low-level locks in C++ \cite{cpplock}.

\subsection{Semaphore}
In this section, we describe the mechanism of the semaphore synchronisation method, as its functionality is replicated within our synthetic dataset. The semaphore is represented by a positive integer variable (also called the count), along with two methods, which we call $up$ and $down$. Whenever a thread $up $'s a semaphore, its count is incremented by 1. Whenever a thread $down $'s a semaphore, either two things can happen depending on the value of the semaphore count. If the value is greater than zero, the $down$ method simply decrements the count by 1 and the program continues execution. If the semaphore is zero, the thread calling the $down$ method momentarily pauses execution until another thread calls the $up$ method, incrementing the counter such that it becomes non-zero. At most one thread can resume its execution time per $up$ operation. An important property of the semaphore is that the $up$ and $down$ operations are both atomic, meaning no two threads can access the shared counter at the same time. The correct pattern of using a semaphore to ensure mutual exclusion to a critical section is shown below \ref{semaphore}.
\begin{lstlisting}[label=semaphore,caption={Mutual exclusion semaphore pattern},frame=tb]
down(semaphore)
critical section...
up(semaphore)
\end{lstlisting}

\bigbreak

There can also be other synchronisation uses of semaphores, one of which is commonly known as the producer-consumer problem \cite{producerconsumerwiki}. In this case, one thread is called the producer, adding an item to a buffer that is shared with the second thread, which is called the consumer, as the latter takes an item from the buffer and processes it. Synchronising the two threads can be done with semaphores as shown below \ref{producerconsumer}. 

\begin{lstlisting}[label=producerconsumer, frame=tb, caption={Producer-consumer semaphore access pattern}]
initialise semaphore count to 0
...
    Thread 1                | Thread 2
    ...                     | ...
    add_item(buffer, item)  | down(semaphore)
    up(semaphore)           | item = remove_item(buffer)
    ...                     | process(item)
    ...                     | ...
\end{lstlisting}

Here we assume that the buffer has unlimited capacity. We show how a semaphore can be used to ensure the consumer always waits for an item to be put in the buffer before it can remove it and process it (should removing an item from an empty buffer be considered erroneous).

\subsection{Concurrency bugs}
\label{subsection:concurrencybugs}
\begin{figure}[!htbp]
    \centering
    \subfloat[Taken from \cite{flanagan2004atomizer}, this showcases an atomicity violation that occurs due to the wrong assumption that the append() function is atomic]{{\includegraphics[width=7.5cm]{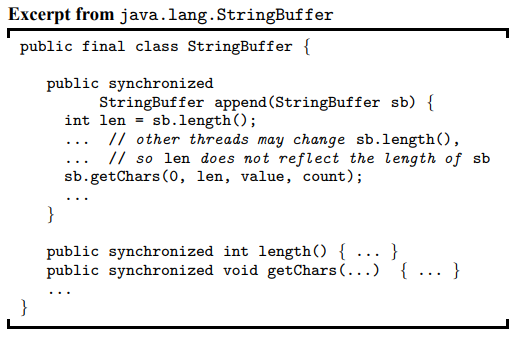} }}%
    \qquad
    \subfloat[Dirty write anomaly \cite{lostupdate}: this specific interleaving leaves the bank in an inconsistent state, with a non-zero difference between the rates of the two accounts]{{\includegraphics[width=6.5cm]{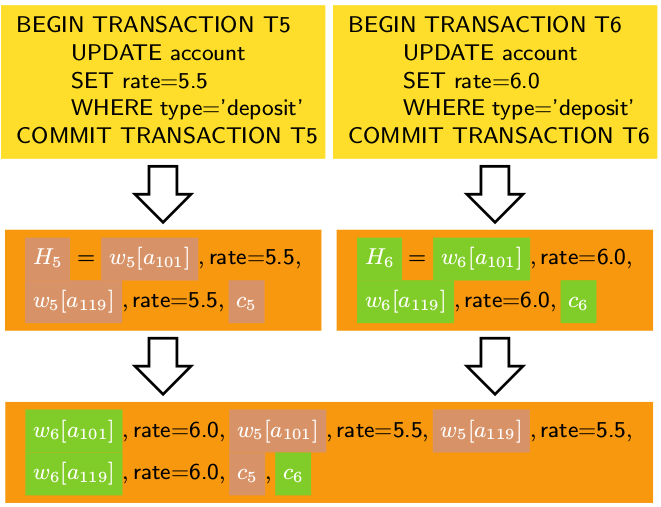} }}%
    \caption{Two more examples of how atomicity violations can lead to undesirable behaviour}%
    \label{fig:atomicity violations}%
    \end{figure}

Concurrency bugs can be attributed to either the misuse of synchronisation mechanisms or the lack thereof. While the reasons behind the misuse of them can be complex and varied, omitting them can be due to the programmer making certain wrong assumptions about the programming language or the program itself. We summarise below four main categories of concurrency bugs depending on what causes them, how they manifest and the negative effect they have on the program.
\begin{itemize}
\item \textit{Atomicity violations}\\
    These occur whenever multiple threads access shared memory at the same time, thus violating atomicity \cite{flanagan2004atomizer} and leading to unintended behaviour. One such unintended behaviour is summarised here \ref{fig:lost_update} but many other anomalies can also occur, depending on the operations and thread interleavings, as exemplified in Figure \ref{fig:atomicity violations}.

\item \textit{Race conditions}\\
    Race conditions are similar in negative effects to atomicity violations \ref{fig:lost_update} but we make the distinction of what causes each one of them. While we consider the main causes of atomicity violations to be wrong assumptions about atomicity or the misuse of synchronisation mechanisms, one main cause of race conditions would be the lack of synchronisation mechanisms due to either simply forgetting or having wrong assumptions about the speed of thread execution. For example, if one were to remove the \lstinline{down(semaphore)} line from our producer-consumer example above \ref{producerconsumer}, you could have a race condition with Thread 1 racing to add an item to the buffer before Thread 2 can attempt to remove an item from an empty buffer.
\item \textit{Deadlocks}\\
    These occur whenever two threads or more are paused waiting for each other to release their respective resources, leading to a circular dependency \cite{deadlockswiki}. Deadlocks usually lead to no progress being made in the program, with the operating system usually having to intervene and break up the cycle.

\item \textit{Livelocks}\\
    Livelocks \cite{deadlockswiki} are very similar to deadlocks. The difference between the two is that in the case of livelocks, instead of the threads being paused, they reset themselves to a previous state before entering the circular dependency. However, if it is the case that they always re-enter the circular dependency, a livelock is formed. They are therefore giving the illusion that the program is making progress when in reality it's not.
\end{itemize}

\begin{lstlisting}[label=verb1, caption={In this simpler example, the deadlock occurs when both threads are waiting at line 6. It could be easily avoided if both threads would follow the same order of $down$ and $up$.},frame=tb]
0: semaphore_1 = 1
1: semaphore_2 = 1
2: ...
3:    Thread 1                | Thread 2
4:    ...                     | ...
5:    down(semaphore_1)       | down(semaphore_2)
6:    down(semaphore_2)       | down(semaphore_1)
7:    critical section...     | critical section...
8:    up(semaphore_2)         | up(semaphore_1)
9:    up(semaphore_1)         | up(semaphore_2)
10:   ...                     | ...
    \end{lstlisting}
    
\newpage

\begin{figure}[!htbp]
    \centering
    \includegraphics[width=15cm]{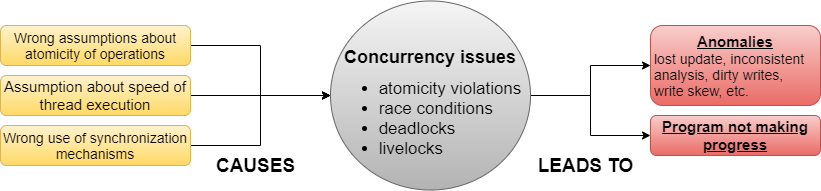}
    \caption{Visual overview of concurrency bugs}
    \label{fig:concurrency_bugs}
\end{figure}

These concurrency issues, what causes them as well as the unintended behaviour they might lead to can be visualised in the diagram at \ref{fig:concurrency_bugs}, which is by no means exhaustive of all possible concurrency problems. \\

\subsection{Current concurrency tools}
As mentioned before, the existing work on concurrency bug detection does not consider all possible traces of concurrent execution, as they grow exponentially. Instead, they are rule-based, with known approaches like \cite{flanagan2004atomizer, wang2005static} expanding on type systems for concurrency like the one shown here \cite{flanagan2003type}. They utilise reduction theorems to condense large blocks of code into individual atomic operations, simplifying the analysis. These theorems are built on top of the type system, with different operations types exhibiting different reduction semantics. For example, the entire code snippet at \ref{semaphore} can be reduced to a single atomic operation, due to the correct use of the semaphore as a synchronisation mechanism.

Other approaches do not utilise type systems but instead maintain monitors of both the shared memory of the program and the synchronisation mechanisms used to access it, like is the case of this paper \cite{savage1997eraser}. There exist trade-offs between all of the mentioned approaches in terms of performance and accuracy. However, as the complexity of concurrent programs increases, it becomes significantly harder for a human to reason about buggy patterns and write tools against them.

\section{Artificial neural networks}
Although the theory behind neural networks has been around for a long time \cite{rosenblatt1961principles}, the recent increases in computational power combined with the big data revolution meant neural networks could finally start outperforming previous AI approaches \cite{mltimeline}. We will not be covering the whole story behind neural networks as it is a common topic with plenty of online resources \cite{nnresources, nnvideo}, but we will cover the important aspects of the neural network variations that are of interest to us for this project.

\subsection{Convolutional neural networks}
CNN's are a subclass of neural networks, theorised in 1980 \cite{cnn} and achieving the best performance on the ImageNet competition in 2012 \cite{krizhevsky2012imagenet}. The convolutional layer has several advantages over its standard, dense counterpart, the most known one being that it is significantly more computationally efficient. However, another advantage is of particular interest to us, which is that of locality. Instead of having a global view of the input features like standard feedforward networks do, the convolutional layers have a local view of the features. In problems with a spatial component (like computer vision, natural language processing or our problem), this allows them to learn features irrespective of their absolute positions in the input feature vector.

\subsection{Long short-term memory neural networks}
LSTM's \cite{hochreiter1997long} are a variation of recurrent neural networks (RNN's) that is of interest to us for their ability to successfully learn sequential data, like in the case of speech recognition \cite{tian2017deep}, financial series prediction \cite{kim2019financial} and natural language processing \cite{young2018recent}. Indeed, our problem of predicting properties from code can be very similar in input feature structure to the problem of natural language processing, hence our interest in RNN's. LSTM's also improve upon the basic RNN architecture by improving upon one important flaw, which is that of the vanishing gradient problem for long sequences \cite{vanishinglstmwiki}. Much like CNN's, RNN's can also learn features mostly irrespective of their absolute positions in the input feature vector, which is what allows them to learn the tasks described above successfully.

\subsection{DeepSets}
Another neural network architecture that is of interest to us is the more recent DeepSets architecture \cite{zaheer2017deep} which looks at modelling input features as permutation invariant, i.e. if the order of the features in the input vector would be changed the machine learning model would still make the same prediction. They formally demonstrate that only a sum over the input features can be permutation invariant and they implement this property in a neural network layer. They demonstrate success with their architecture in several applications, the one closest to our problem being the digit sum prediction \cite{digitsum}, where they achieve better convergence than LSTM's. Permutation invariant modelling of features is of particular interest due to the concurrent aspect of our problem. Consider the following: if multiple threads run in parallel, then at runtime there are exponentially many possible global orderings between all operations. However, we want a single model prediction to account for all possible orderings, since the actual orderings of operations can only be observed at runtime, and exploring all orderings is unfeasible as discussed before. Ideally, a model should classify a program as buggy if no possible ordering/interleaving would lead to unintended program behaviour, or similarly if all possible orderings are in fact equivalent in terms of program output.

\section{Machine learning model interpretability}
\label{section:2.3}
One of the biggest issues in machine learning nowadays is the black-box nature of certain models, with more complex models achieving better accuracy but allowing for less interpretation \cite{lundberg2017unified} (neural networks being the most notable example of this). Among all interpretation methods, we can distinguish two important and mutually inclusive approaches to understanding the learning mechanisms of a model. \\
 
 The first approach seeks to understand how input features influence the model prediction, both in terms of absolute importance, as well as the direction towards which they influence the prediction. This method of interpretability works locally around each prediction, and first of all involves creating a secondary dataset around each sample through feature variations of the original sample. Afterwards, a very simple model is trained to match the predictions of the original model on this secondary dataset. The weights of the simpler model indicate the importance of features as well the direction towards which they influence the prediction \cite{6883}. Tools like LIME \cite{ribeiro2016should} and SHAP (SHapley Additive exPlanations) \cite{lundberg2017unified} allow for this sort of analysis with minimal effort. \\
 
 The second interpretability approach is the one we are using throughout this project. This approach attempts to measure the generalising capabilities of a model by selectively filtering the train and test distributions. The motivation for this sort of interpretation is that machine learning model accuracy can be significantly impacted when the underlying train and test distributions are different \cite{lakshminarayanan2017simple}, when adversarial networks are employed \cite{goodfellow2014explaining, niven2019probing} or when there simply are not enough samples to train on \cite{wang2018deep}. Although there already exist tools for the probing of models \cite{wexler2019if}, we develop our own dataset as well as filtering methods to have fine-grained control. Indeed, the approach of using synthetic datasets to test/validate properties about neural networks is not unheard of \cite{merrill2020formal}.

\newpage
\chapter{Synthetic dataset}

Although in the beginning we had an idea of the sort of neural network properties we wanted to test, our experimenting process is undoubtedly an iterative one. With each set of experiments that we run, we improve our overall understanding of how some specific model types perform better than others on certain inputs. This, in turn, allows us to think of more scenarios to test for, as well as add more complexity to our synthetic dataset as a result. 

\begin{figure}[!htbp]
    \centering
    \includegraphics[width=7.5cm]{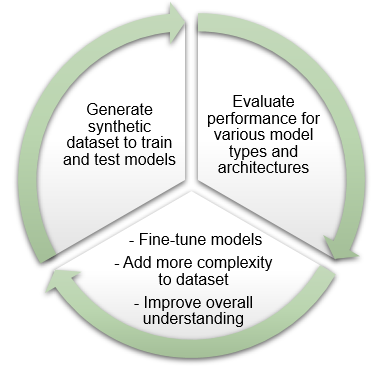}
    \caption{Visual summary of our process - iterating on experiments is crucial to improving our understanding of the task and context}
    \label{fig:pipeline}
\end{figure}

\section{Python}
Our choice of language for generating and manipulating the synthetic dataset as well as developing the machine learning models is Python \cite{10.5555/1593511}. Not only is it accessible and fast to develop making it perfect for experimenting, but it also supports a vast range of machine learning libraries, including but not limited to Keras \cite{chollet2015keras}, PyTorch \cite{paszke2017automatic}, Scikit-learn \cite{scikit-learn}. To this end, we also utilise the Jupyter Notebook \cite{PER-GRA:2007}, which further facilitates development.

\newpage

\section{Grammar for synthetic data generation}

More concretely, each sample in our synthetic dataset consists of arrays of characters (i.e. strings), where each character can be thought of as a token that represents a program operation. This dataset is meant to model simple patterns we might see in (concurrent) programs, as well as certain patterns that may be harder for machine learning models to capture due to fundamental limitations. We define "concurrent" operations that either access the shared memory or synchronise access to shared memory. We also define "no-op" operations that do not affect the shared memory in any manner but instead act as noise. There are also two simulated objects to speak of, namely a shared variable and a synchronisation mechanism. The use of these operations, as well as the interaction between them, dictates whether a generated sample is considered buggy or not, based on some underlying rules that we define. \\

To thoroughly test out the different variations of machine learning models and architectures, we generate all possible combinations of operations at all possible positions in the program. At the same time, we only utilise a certain subset for training the models. How we select this subset is indeed very important, since we can test the models' abilities to generalise by selecting which samples a model can see during the training stage, and which samples it can see during the validation/test stages (more details regarding the filters in the 4th chapter \ref{chapter:4}). \\

We generate 3 strings of characters meant to model 3 concurrent Functions being run in parallel. The first Function type does not access the shared memory and thus has no impact on whether a sample is labelled as buggy, while the second Function attempts to write/initialise a variable that is then read by the third Function. We deem a program to be buggy if there exists any possibility that the read in the third Function occurs before the initialisation in the second Function. \\

We cover each operation, Function and underlying rule as well as their relevance to the truth label of a sample in the upcoming sections. We also provide visual summaries of our grammar under the form of tables at \ref{fig:tokens}, \ref{fig:token examples}, \ref{fig:bnf} as well as actual examples of generated samples at \ref{samplesexamples}. We also motivate our choice of operations, Functions and underlying rules in the upcoming sections. 
For the moment, a concrete comparison could be drawn to our previous producer-consumer example \ref{producerconsumer}: a buggy sample in our case (i.e. reading before writing) is similar to attempting to remove an item from an empty buffer. In the producer-consumer example that erroneous behaviour is avoided through the use of a synchronisation mechanism (typically a semaphore), which is something we also replicate in our toy use case.

\subsection{Function 1 (no-op Function)}
    The first type of Function does not affect the global state at all. It is a random regex of the form '(,.\_)+' where each symbol (, . \_) is to be interpreted as an operation that does not affect the global state of the program. For instance, .,\_.\_,...\_.\_.\_. might be such a generated string of characters that corresponds to a particular function whose execution is '.' followed by ',' and then '\_' and so on. Much like the operations it contains, the purpose of this Function is to be noise which the models should learn to ignore.

\subsection{Function 2 (writing Function)}
    The second type of Function affects the global state of the program by writing/initialising a shared variable.
    It can do either of the following 3 different things or nothing at all: 
    
    \begin{itemize}
    \item Just writes/initialises the variable (we denote this by 'w')
    \item Writes/initialises the variable and ups a semaphore afterwards. This represents a correct use of the synchronisation mechanism. We denote this by 'wu'
    \item Ups the semaphore and then writes/initialises the variable. This represents a misuse of synchronisation as the orders are wrong. We denote this by 'uw'
    \end{itemize}
    
Function 2 can also do any number of non-state-affecting operations before, in-between and after the state-affecting operations. One such pattern could be '.\_,w\_u.,..'. 
    
\subsection{Function 3 (reading Function)}
    Finally, the third type of Function can access the global variable to which the second Function writes to. However, accessing it without Function 2 writing to it first would constitute an erroneous program. The third type has the following actions:
         \begin{itemize}
            \item Just reads the variable, denoted by 'r'
            \item Checks if the value was written to before. Either reads the shared variable if yes or reads a default value if no. We denote this by 'cr'.
            \item Downs the semaphore and then reads the variable. This represents a correct use of a synchronisation mechanism. We denote this by 'dr'.
            \end{itemize}
    
Hence one possible string generated for this Function can be '..,\_\_r.'

\subsection{Illustrating the grammar}

\begin{table}[!htbp]
    \centering
    \includegraphics[width=15cm]{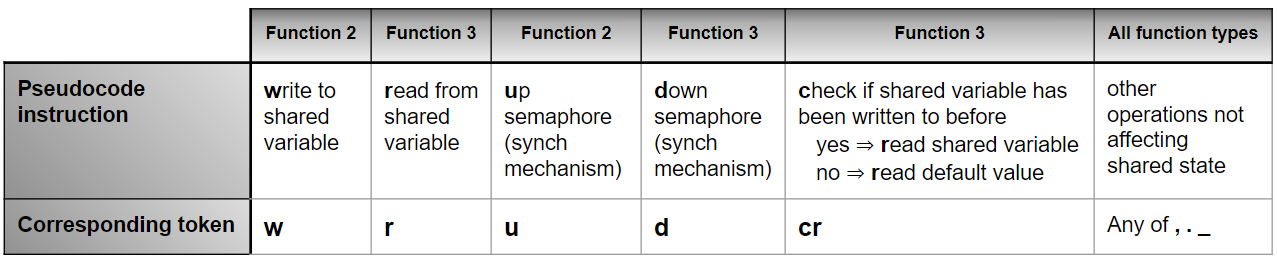}
    \caption{Table that summarises the correspondence between pseudocode instructions and the tokens that model them}
    \label{fig:tokens}
\end{table}

\begin{table}[!htbp]
    \centering
    \includegraphics[width=15cm]{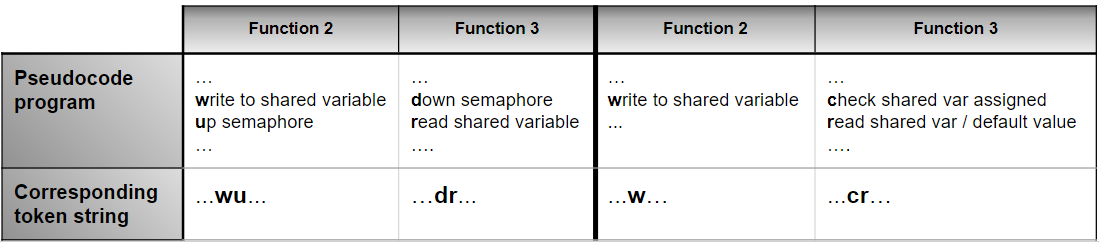}
    \caption{Example correspondence between 2 pseudocode programs and their simulated counterparts}
    \label{fig:token examples}
\end{table}

\newpage
\begin{table}[!htbp]
    \centering
    \includegraphics[width=15cm]{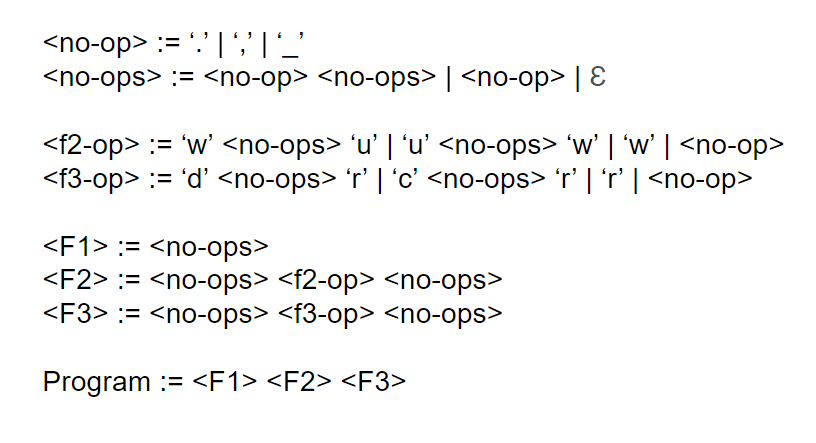}
    \caption{Formal BNF grammar for our synthetic dataset}
    \label{fig:bnf}
\end{table}

\bigbreak

\begin{lstlisting}[label=samplesexamples, caption={Example: 20 random samples, under the form of arrays of strings}, frame=tb, language=python]
#   Function 1          Function 2          Function 3        Label
[[',,..,__._,.,._.,', '______.,_.uw._.,', '___..,__.__dr_._', True],
 [',___,,._,,.__.,.', '__._,__w__._,,,,', '_,.,,,.,_.dr_...', True],
 [',...__..,.,,,_..', '.,___.__.wu___.,', '._cr._..__,,,.,.', False],
 ['.,,._.._.,_,,,..', '_..__,,,.,.__.,,', ',_,._._,_,.r,,__', True],
 ['.__.,,.,,_.,___,', '_._,,,._..._,_,,', '._.,_dr.,,.._,__', True],
 [',,__,,_.___...,_', '_,__,_,_,,_._..,', '_,.,,cr.._,.._,.', False],
 ['.._...,.,._,,.,,', ',._,,.,,,.w_._.,', ',,.,,__.,.__,__,', False],
 ['._.,..._.._...._', '_,_...__.,,.wu,_', '_.._.,.._._,_,._', False],
 ['.._._,,_._,....,', '_....__,.,uw,_._', ',.,_,,.._,__r,,,', True],
 ['.,_..,,._,__.,,.', ',___,_.._,_,w.,_', ',,._,,..__,cr__.', False],
 [',_._,___.,.,,,.,', ',.,_.__.__,,,...', '.,,..___.__,,.cr', False],
 ['_______,.,_.,._,', '.,_._,,__,.,.wu,', ',_.__._,,,.,__,,', False],
 ['__..,.._.,__,.__', '.___,_.w_._,_...', ',_,_,.,_..,,____', False],
 ['.._.,.,,._.._..,', ',..,_w..,.,,.,_.', ',_..,,._r__,,_,_', True],
 ['_____.___._,..,,', '.,___,._,.,w_,._', ',_.,r..,_,._,_,,', True],
 [',.,,_.,,,__.,__.', ',.,,,,.,u,,w.._,', 'd,_r_,_....,,._,', True],
 ['._.,,_,,.,._,_,.', '.__.__._uw..,,,,', '_.__,,_.dr_,,.._', True],
 ['.,__,.,.....,.,.', '..__uw_.___,___,', '__,..._,.r._,.,_', True],
 ['_._____.._.._,,,', '.,__wu._,,,.____', '_,___cr,_.,_.__.', False],
 ['...._..,_.,.._,,', '._w..u_.__..,.,.', '__._,_.,._,,d.,r', False]]
\end{lstlisting}

 \newpage
 
 \section{Underlying concurrency patterns}
 Once we defined a grammar corresponding to a simple concurrent program, we can then define some rules that decide whether a combination of the three randomly generated Functions corresponds to a buggy (\xmark) program or not (\cmark). \\
 
 Therefore, like in our producer-consumer example, if Function 3 attempts to read the shared variable without any synchronisation mechanism or checks, the program has a potential race condition. Thus, any combination of Functions where the third Function contains only the 'r' operation is buggy (\xmark) (see 2nd column of below table \ref{fig:labels}). \\
 
 Not reading the shared variable (i.e. absence of 'r' character in Function 3), or checking before reading the shared variable (i.e. presence of 'cr' in Function 3) both constitute non-buggy programs (\cmark) (see 1st and 3rd columns of below table \ref{fig:labels}). \\
 
 The only case where Function 2's operations are relevant to the truth label of the program is whenever Function 3 correctly uses the synchronisation mechanism (i.e. presence of 'dr' in Function 3, 4th column of below table \ref{fig:labels})). In this case, the program is valid (\cmark) only if Function 2 must also correctly use the semaphore by first writing to the shared variable and then upping the semaphore, corresponding to 'wu'. The assumption here is that the semaphore count is initialised to 0, hence Function 3 can only proceed with reading the shared variable after Function 2 has incremented the semaphore count. If no semaphore is used, the program is considered buggy (\xmark) as Function 3 would become stuck on downing the semaphore. If the semaphore is misused as is the case when you up the semaphore before writing to the variable ('uw'), the program is also considered buggy (\xmark). This is because 'dr' in the third Function could interleave between 'u' and 'w' (i.e. global execution trace of $u \xrightarrow{} d \xrightarrow{} r \xrightarrow{} w$). \\

\begin{table}[!htbp]
    \centering
    \includegraphics[width=12cm]{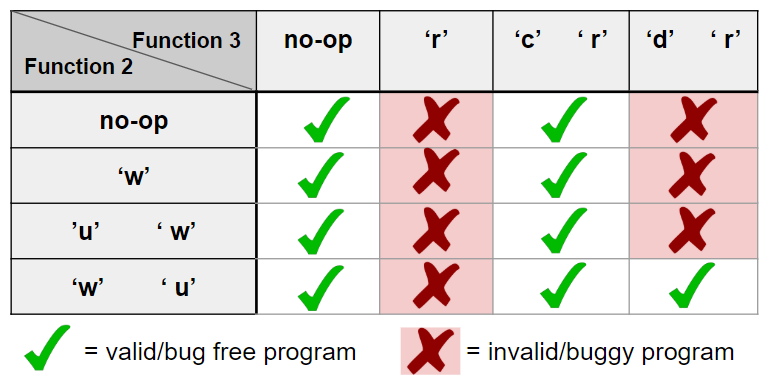}
    \caption{Table summarizing how the interaction between Function 2 and Function 3 determines the label of a 'program' (green for valid program (\cmark), red for buggy/invalid program (\xmark))}
    \label{fig:labels}
\end{table}

\newpage

\section{Reasoning behind our choice of grammar and rules}
As mentioned before, we somewhat inspire our grammar and rules from the producer-consumer example \ref{producerconsumer}. However, the producer-consumer pattern is not something one might encounter as often as some of the other concurrency patterns (for e.g., the mutual exclusion example \ref{semaphore}). Nevertheless, we believe that our grammar offers better model interpretability due to a clear separation of operations between the two Functions. If both Functions were to have the same or similar set of operations as would be the case if we were to model mutual exclusion \ref{semaphore}, we could not test to such an extent the model properties as we will show in the Evaluation chapter. 

Additionally, depending on the combination of operations from Function 2 and Function 3, one might draw a comparison to the standard concurrency bugs discussed during the background section \ref{subsection:concurrencybugs}. Not using any synchronisation mechanism in Function 3 causes a race condition, not using any synchronisation mechanism in Function 2 causes a deadlock when Function 3 downs the semaphore and misusing the synchronisation mechanism in Function 2 constitutes an atomicity violation.

\subsection{Complexity of rules}
We design the underlying rules such that there are two levels of complexity when it comes to learning the patterns. The first level is the easiest one as models only have to consider the operations of Function 3. However, the second level of complexity requires that models also consider the operations of Function 2 as well as the interaction between Function 2 and Function 3. Models can only successfully learn the complex rules if they succeed with learning the simpler rules first. \\

\bigbreak

\begin{minipage}{0.5\textwidth}
\includegraphics[width=\linewidth]{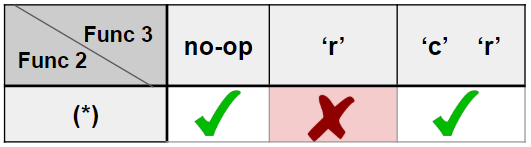}
\captionof{table}{Simple rules}
\label{fig:simple rule}
\end{minipage}%
\hfill%
\begin{minipage}{0.45\textwidth}\raggedleft
\hspace*{2cm}
\vspace*{-1cm}

\bigbreak

\begin{itemize}
\item Out of 16 possible cases (4 cases each for Function 2 \& 3), 12 can be correctly classified by only considering Function 3.

\item We test only for the capability of models to learn the inclusion of operations 'c' and 'r'

\end{itemize}
\end{minipage}

\bigbreak
\bigbreak
\bigbreak

\begin{minipage}{0.3\textwidth}
\includegraphics[width=\linewidth]{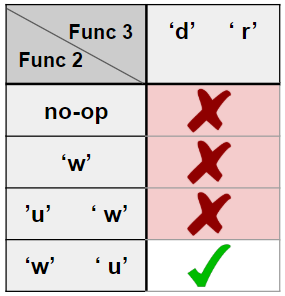}
\captionof{table}{Complex rules}
\label{fig:complex rule}
\end{minipage}%
\hfill%
\begin{minipage}{0.6\textwidth}\raggedleft
\hspace*{1cm}
\vspace*{-0.5cm}

\begin{itemize}
\item In 4 out of 16 cases models have to consider the operations of both Functions

\item Not only is it not enough for a model to learn the inclusion of both 'u' and 'w', but the relative ordering of the two is actually significant: 'uw' is considered buggy (\xmark) while 'wu' is considered valid (\cmark)

\item Models that are not able to learn these complex patterns predict every sample as buggy (\xmark) when 'dr' is present in Function 3, as that gives the best compromise 

\item We also notice empirically \ref{patternresults} that the 'wu $|$ dr' case has posed the biggest challenge for our models

\end{itemize}
\end{minipage}

\bigbreak

\newpage

\section{Dataset parameters}
Through the use of a flexible API, we can control various parameters of the synthetic data generation. These include:
\begin{itemize}
    \item \textbf{Controlling the number of operations that can appear in each sample}
    
        We can control the number of operations per Function in each sample. For example, the samples at \ref{samplesexamples} are generated with 10 operations per sample. Varying the number of operations per Function also lets us control how many total samples can be generated, as the number of possible operation combinations grows quadratically with the Function size. By varying the number of operations between samples of the same dataset, we could also test for robustness of programs to size variance of programs.
        
    \item \textbf{How many no-op operations appear in-between state-affecting operations} 
    
        Under our rules, the variance in the number of no-op operations between samples (e.g. "wu", "w.u" and "w..u" have inter-operational distances of 0, 1, and 2 respectively) should not affect the truth label of a sample. Therefore the models should also be able to generalise based on this. This parameter is closely tied with Filter \#4 as described in the next chapter.
        
    \item \textbf{A specific order in which the samples appear in the dataset, ordered by the particular patterns they contain and the positions at which they appear} 
    
        Generating patterns in a specific order is useful not only for visualising but also for being able to clearly separate samples based on the positions of operations (such a separation is illustrated in the next chapter \ref{table:filter2 example}). This parameter is closely tied with Filter \#2 as described in the next chapter.
\end{itemize}

\newpage

\chapter{Dataset Filters}
\label{chapter:4}

As discussed in \ref{section:2.3}, one method of interpreting model capabilities is through the careful separation of train and test samples, which will also be our approach and theme of this section. Initially, the only method of separating train and test was through the random subsampling of the full dataset. However, as our understanding of the problem increased, we could add more Filters and test for more model properties. We have thus far developed four such Filters, with future work including the development of others. We discuss these Filters in the following sections, but it is always useful to refer to the examples table \ref{table:filters}.

\begin{table}[!htbp]
    \centering
    \includegraphics[width=15cm]{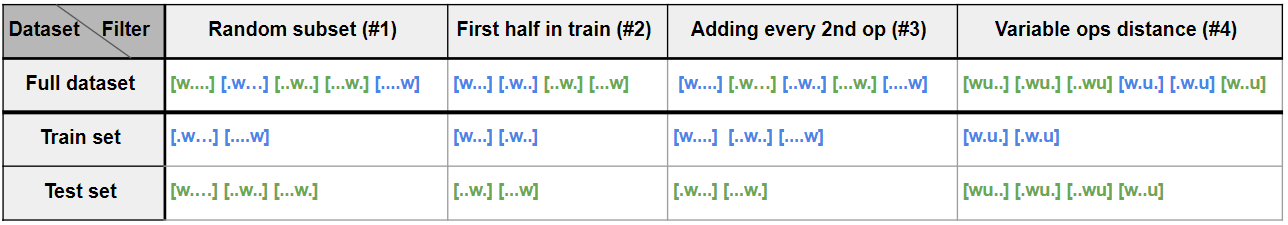}
    \caption{Table of examples showcasing how the individual Filters separate a full dataset into train and test}
    \label{table:filters}
\end{table}

\section{Filter \#1 - random subset} 
\textbf{Random separation of the full dataset based on a given train-to-test ratio parameter} 

This is the most basic of Filters and is supposed to reflect the real-life randomness of programs while also serving as a simple baseline to compare performance on other Filters.

\section{Filter \#2 - first half only} 
 \textbf{Separation of samples where meaningful operations appear only in the first half of a program vs. samples where meaningful operations appear only in the second half of a program} 
 
This is used to test a model's robustness to variance in the position of operations. Even if a model can correctly classify a pattern that only occurs in the first half of a program during training, it should be able to generalise to correctly classify the same pattern if it were to occur in the second half of a program during testing. 

The table at \ref{table:filters} simplifies the effect of this Filter as it only displays the separation of a single Function. In reality, the separation of both Functions is required. The samples of the full dataset are generated in the order displayed at \ref{table:filter2 example}. They are also separated such that only samples whose Function 2 and Function 3 operations are positioned in the first half can appear in the train dataset (and equivalently second half for the test dataset). This leads to the discarding of a subset of samples, which is best illustrated in this table \ref{table:filter2 example}. The reasoning for discarding those samples from test is that we get a clearer picture of models' robustness to position variance during testing. \\

\begin{table}[!htbp]
    \centering
    \includegraphics[width=15cm]{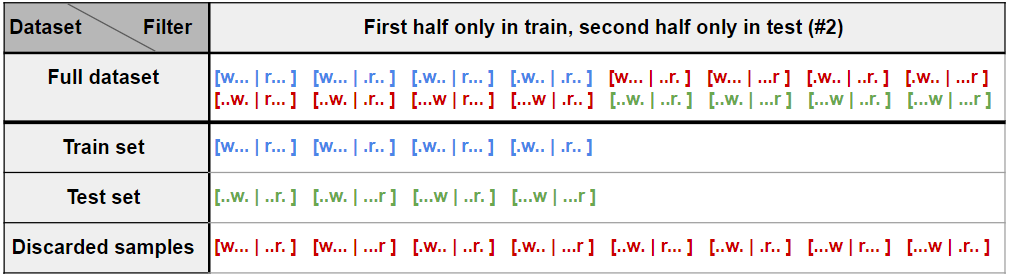}
    \caption{Filter 2 separation taking into account positions of operations in both Functions}
    \label{table:filter2 example}
\end{table}

In essence, the separation for two Functions is done by assigning the first fourth of samples to train, assigning the last fourth to test, and discarding the rest.
 
 This Filter can also be generalised to separate the first $1/n$ ordered samples in train, and the last $(n-1)/n$ interval in test, although extra care needs to be taken, especially when combining it with the other Filters.
 
\section{Filter \#3 - skipping every n samples}
\label{section:filter3}
\textbf{Only considering every n'th ordered sample as part of the training process = skipping n-1 samples every time before adding one to the training set}

Similar in purpose to the first Filter, but this time the separation is deterministic so as to reflect bias in the train test vs test set. Depending on the actual value of n, an interesting effect can be noticed, as shown below. \\

\begin{lstlisting}[label=filter3, caption={First 10 samples from the training set after applying the third filter}, frame=tb, basicstyle=\small]
             Filter skip size = 5     |      Filter skip size = 6
              Func 2      Func 3      |       Func 2      Func 3
         ['w___._.,,_' 'cr,.,_,__.']  |  ['w___._.,,_' 'cr,.,_,__.']
         ['w,__._,,.,' '.,.__cr_,_']  |  ['w__.,,,.,,' '.._.._cr,_']
         ['_w..,,..._' '.cr,.,__.,']  |  [',w_._,,.__' ',,.cr...__'] 
         ['.w_.,__,,,' '_,.__.cr._']  |  ['_,w,._,..,' 'cr,,._,__.']
         ['..w,___.,_' '.,cr_,__,.']  |  ['_.w__,_..,' '.,..._cr__']
         ['_,w_.,,._,' ',....,.cr_']  |  ['.._w,..,_.' '..,cr..,._']
         ['.._w,..,_.' '..,cr..,._']  |  ['.,__w_,,.,' 'cr_,,._._.']
         [',.,w__.._,' ',,,._,_.cr']  |  [',..,w,.,._' ',,,.__cr._']
         ['___.w..._.' '._.,cr,.,,']  |  ['__...w___,' '...cr.,,._']
\end{lstlisting} 

\bigbreak

There is an interesting difference between the possible positions of the 'cr' operation in the Function 3 code depending on the skip size used. In the left example, 'cr' can appear at any position in the program. However, on the right example, 'cr' can only appear at certain positions, namely 0, 6 and 3.  Therefore, setting the skip size parameter to equal 6 would mean the 3rd Filter would also test for robustness to position variance in code (since the models can see 'cr' at only 3 positions during train out of many more possible positions). 

\newpage

\section{Filter \#4 - variable inter-operational distance} \textbf{Varying the number of no-op operations between meaningful operations = separating samples in train and test based on the distance between meaningful operations}

For example, the train set can contain samples with an inter-operational distance of 1 (e.g. 'w.u'), while the test set can contain samples with an inter-operational distance of 2 or 0 (e.g. 'w..u' and 'wu' respectively). The reasoning behind adding this Filter was that we expected to see a big difference between CNN's and LSTM's, with the latter outperforming the former. We show in the results section that this expectation was indeed met.

\section{Combining Filters}
\label{combiningfilters}
Not only can we use these 4 Filters on their own, but due to the flexible design of the API we can have any combination of these 4, as long as the orders in which the Filters are applied is respected. This is useful because we can test multiple desirable properties (e.g. robustness to variance in operation position, program size, placement of two meaningful operations in relation with each other) by combining different Filters (e.g. combining Filters 2 and 4 respectively). In the given example of combining Filters 2 and 4, the combination also makes sense empirically, as the accuracies for the models in the combined case is an average of the accuracies in the individual Filters \ref{combiningfiltersresults}.

\subsection{Filter orders}
In order to have any combination of the 4 Filters, the following order must be respected:
$$ Filter 2 \xrightarrow{} Filter 3 \xrightarrow{} Filter 1 \iff Filter 4$$

Filter 2 must be applied first as it is highly dependant on the original order the samples were generated in (see \ref{table:filter2 example}). Neither Filter 1 nor Filter 4 care about the ordering of the original dataset, so either can be applied last. However, Filter 3 needs to be applied before Filter 1, as Filter 1 implies shuffling, which ruins the determinism of Filter 3.

\section{Validation set}

We consider the validation set to be part of the testing set, as we use the validation accuracy metric for early stopping so as to prevent overfitting.
Therefore, we allocate a random subset from the test set to the validation set (the split is usually 1:3 validation to test).

\newpage

\chapter{Machine learning models}
\begin{figure}[!htbp]
    \centering
    \includegraphics[scale=0.3]{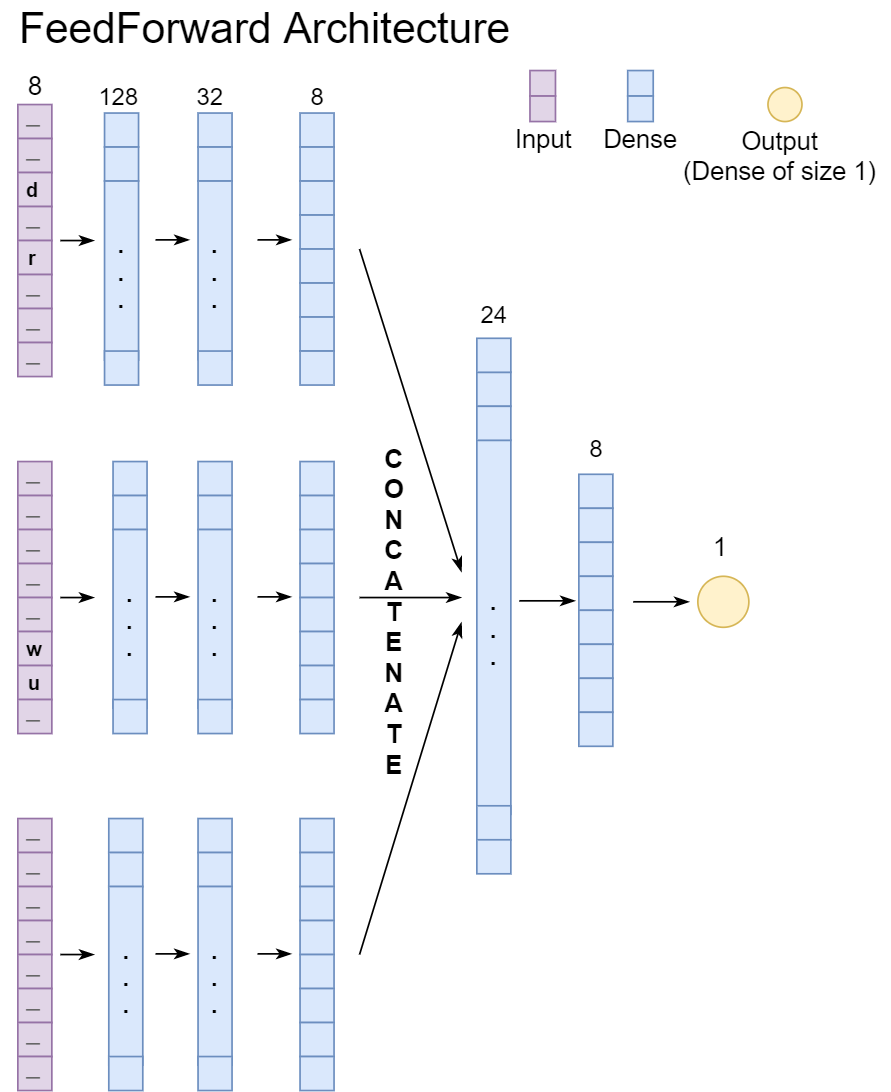}
    \caption{Model architecture for the Feedforward type - only dense layers are used}
    \label{fig:feedforward}
\end{figure}

Because the three Function types are meant to model concurrent code being run in parallel, we need to also separate the input feature vector for each Function. As best illustrated in the overview diagram of our most basic architecture below \ref{fig:feedforward}, the three separate inputs are each linked to three separate series of hidden layers. The separation of the input feature vector for each Function helps the models focus on learning each Function logic irrespective of the other Functions. Then, the three streams are concatenated under one single dense layer, which is then followed by one or more dense layers. This then allows the models to also learn the global interaction between the Functions. \\

We recognise that this approach might be unfeasible for real-life programs due to the high variance in the number of parallel sections for each program (as opposed to our fixed number of three parallel Functions). However, modelling such input is a non-trivial task not within the scope of this project, but future work could include either employing graph neural networks to circumvent this problem or utilising padding or some other mechanism of normalising the input dimensions. \\

The architectures of the different model types are very similar, with most of the differences coming from the special layers (i.e. CNN/LSTM/DeepSets) as well as the number of dense layers or the number of neurons per dense or special layer. We provide illustrative but comprehensive diagrams of the fine-tuned model architectures that were used to obtain our results. From the diagrams we only abstract away the process of one-hot encoding the input.

\newpage

\section{Feedforward}
This model architecture is composed of only dense layers and is intended to serve more as a baseline on which more complex models can further improve on. The first learning experiments on the dataset were done using this model type, and at around the same time we've decided to break down the input feature vector into three parts, as doing so would result in better accuracy. \\

\newpage

\section{DeepSets}
We experiment with two variations of the DeepSets architecture \cite{zaheer2017deep}, each one seeking to employ the paper's result which dictates that only a summing function can truly be permutation invariant.

\subsection{DeepSets variation \#1}
\begin{figure}[!htbp]
    \centering
    \includegraphics[scale=0.3]{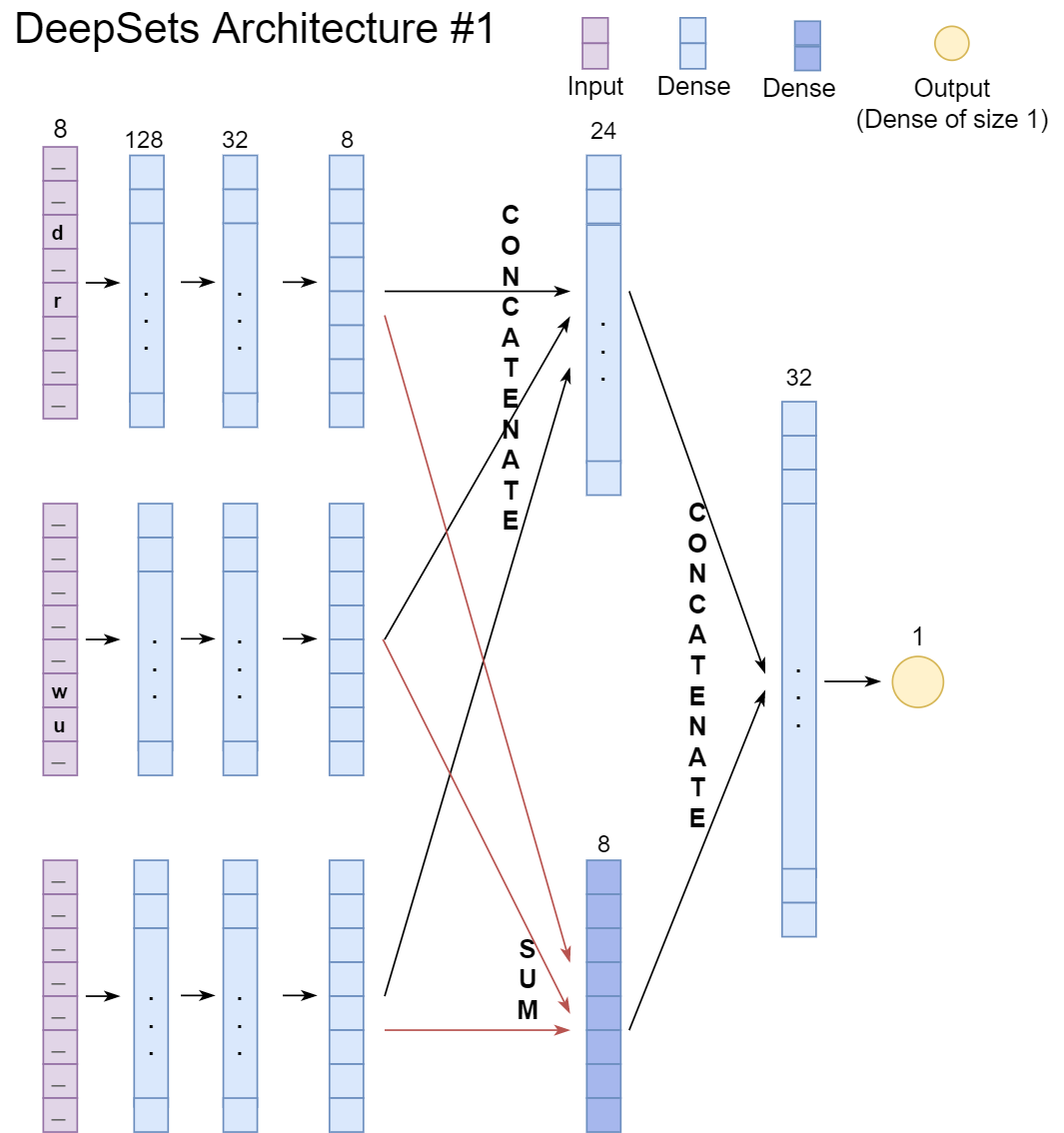}
    \caption{DeepSets variation \#1 architecture, very similar to Feedforward \ref{fig:feedforward}. The single dark blue layer signifies an element-wise summing of the previous 3 light blue layers}
    \label{fig:deepsets1}
\end{figure}

This is a close adaptation of the architecture used in the DeepSets DigitSum experiment \cite{digitsum}. The task of DigitSum is predicting the sum of at most 10 digits, where the representation for each digit is a 28 by 28 matrix of pixel values (like in the MNIST dataset \cite{lecun2010mnist}). The input is separated for each digit, followed by  
10 series of dense layers connected to each digit input matrix, much like in our architecture. The last 10 dense layers are of size 30, which are then joined by performing element-wise summing on the 10 dense layers, resulting in a single vector of size 30. We follow a very close architecture, as shown below \ref{fig:deepsets1}.

However, one difference is that we also add another dense layer in parallel with the DeepSets layer. During early experimentation, we've found that the additional dense layer helps the augmented model perform at least as good as the one without. For thoroughness, future work also includes getting experimental results for the unaugmented model. \\

\newpage

\subsection{DeepSets variation \#2}
This 2nd variation of DeepSets is meant to utilise the permutation invariant summing layer right at the beginning. Much like a simple linear regression model, a weighted sum is performed over the input. The initial purpose of this 2nd variation was to show robustness to the position variance of code while the expectation was that the relative ordering of operations would pose problems. However, running our experiments, we've found that this DeepSets variation was unable to learn anything, even on the train data. 

Future work includes presenting the input in a different manner (e.g. bag of words), such that the weights of the first layer themselves are not dependant on the absolute position of an operation. Although we excluded this particular model type from the discussion of our results, the original performance can still be found under the appendix. \\

\begin{figure}[!htbp]
    \centering
    \includegraphics[scale=0.3]{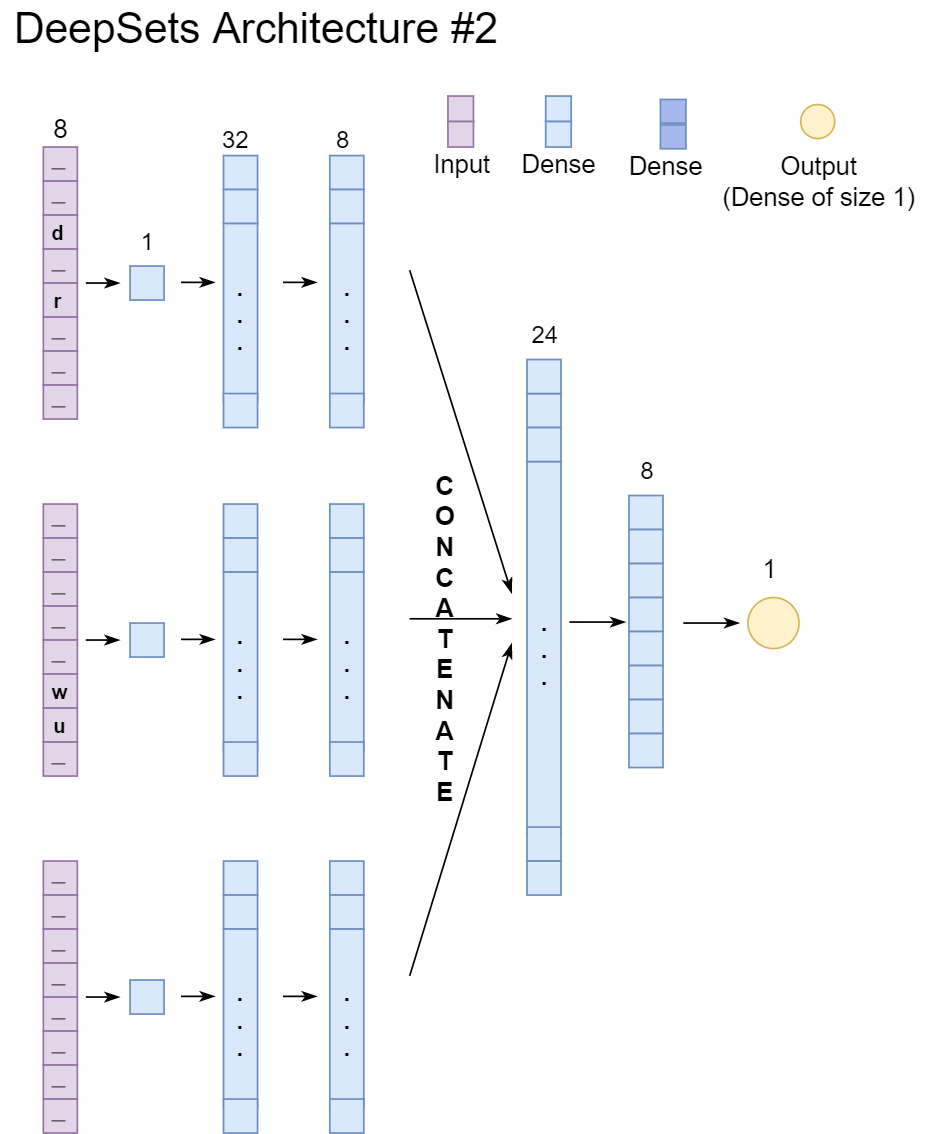}
    \caption{DeepSets variation \#2 architecture, similar to Feedforward \ref{fig:feedforward}}
    \label{fig:deepsets2}
\end{figure}

\newpage

\section{CNN}
Chronologically, the CNN was the third model type we implemented, after Feedforward and DeepSets. It was also the first model type that could achieve near-perfect accuracy when only training on a small subset (10\%) of the full dataset. Below we provide an architectural diagram of the CNN, and in the next pages we delve into the capabilities and limits behind the convolutional filters and why it would be reasonable or not to expect CNN's to classify various patterns correctly. \\

\begin{figure}[!htbp]
    \vspace*{0.2cm}
    \centering
    \includegraphics[scale=0.29]{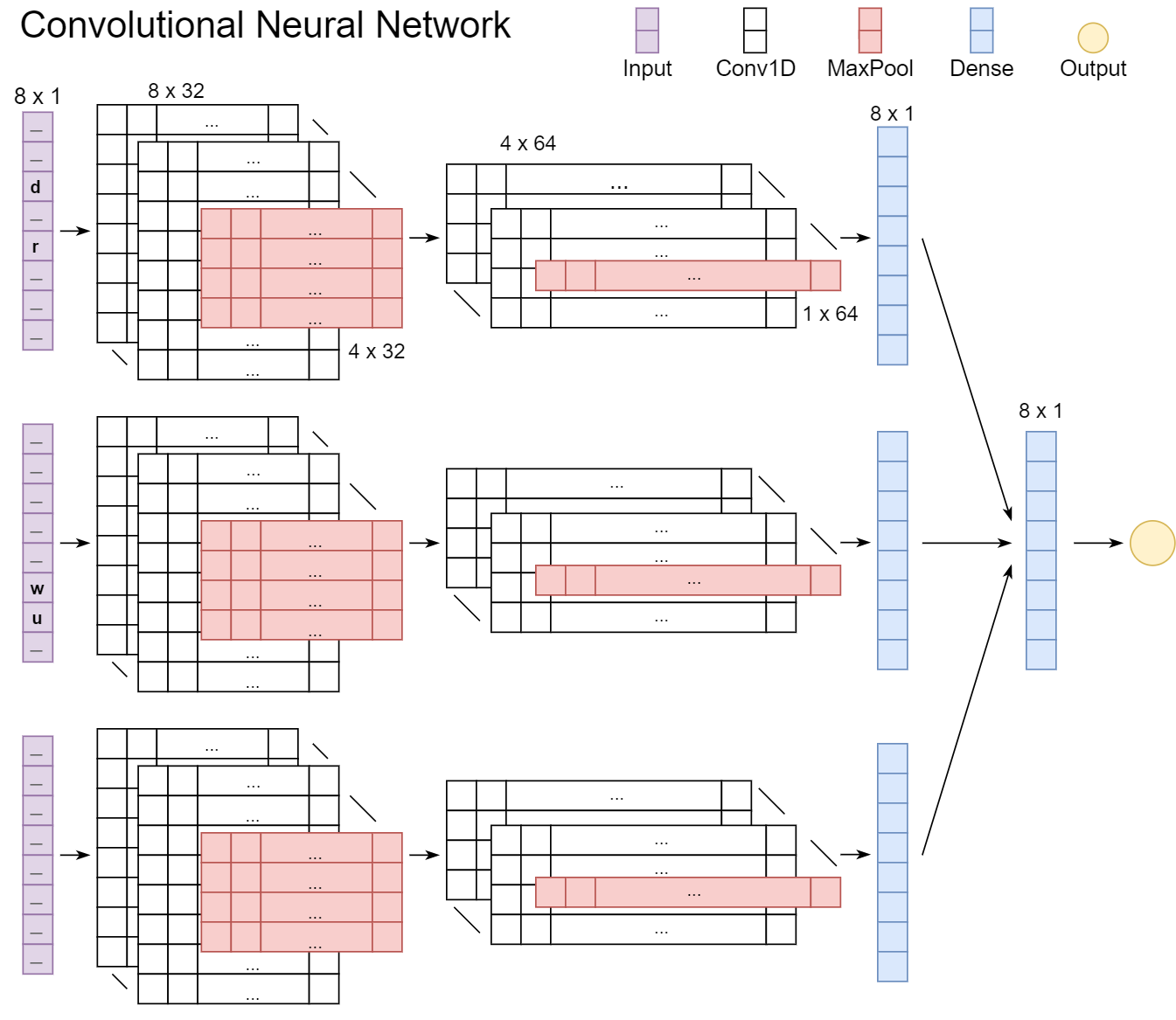}
    \caption{Overview of our fine-tuned CNN architecture, similar to VGGNet \cite{simonyan2014very} in the sense of progressively narrowing the input dimension but increasing the convolutional filter dimension}
    \label{fig:cnn}
\end{figure} 

Since we believe that reasoning about CNN performance on our dataset is particularly difficult, within the next four subsections, we explore how CNN's \textit{might} work behind the scenes to classify patterns successfully. We give examples of how CNN's \textit{might} train their weights to classify samples correctly. We do not claim that empirically CNN's train their weights as shown in the examples. However, we show it is well within their capability to do so and by the simplicity of the examples, we argue that it is also something reasonable to achieve. Future work includes looking in more detail at the actual filter weights to either validate our examples or provide new theoretical insight into how CNN's filter weights affect the classification of patterns. \\
\label{layerweights}

It is also worth mentioning that in theory, CNN's are able to successfully learn all patterns with a small number of filters and layers. In practice, however, a larger number of filters and layers helps them converge to that point by providing the model with a larger pool of potentially useful weights.

\newpage

\subsection{Convolutional filters}
\label{cnn}
Within this subsection, we define notations that we are using to illustrate CNN capabilities. First, we provide an example to remind of the format that one-hot encoded input takes.\\

\bigbreak

\begin{minipage}{0.4\textwidth}
\includegraphics[width=\linewidth]{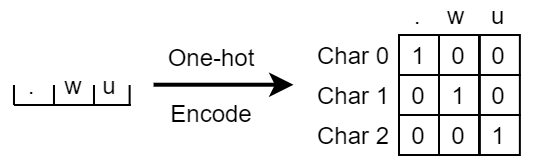}
\captionof{figure}{One-hot encoding the ".wu" string of size 3 results in a binary array of dimensions $3x3$ (which can also be flattened to a single dimension of 9)}
\label{fig:cnn filter 1}
\end{minipage}%
\hfill%
\begin{minipage}{0.5\textwidth}\raggedleft
\hspace*{2.5cm}
\vspace*{-0cm}

In the array on the left, we have an entry of 1 at positions $i, j$ whenever the character corresponding to the j'th column is found at the i'th position in the original input. Otherwise, everything else is set to 0. One-hot encoded arrays can also be flattened and \textit{are} flattened in our code (in this example it would be flattened to the vector $[100010001])$

\end{minipage}
\bigbreak
\bigbreak

Below we show how convolutional filters can be applied to the input to produce a binary output indicating the inclusion of a particular operation. Whenever the convolutional filter "matches" the input, the output is 1, otherwise the output is 0. As the one-hot encoded input can be confusing and cumbersome to visualise, we abstract it away. \\

\begin{figure}[!htbp]
    \centering
    \includegraphics[width=15cm]{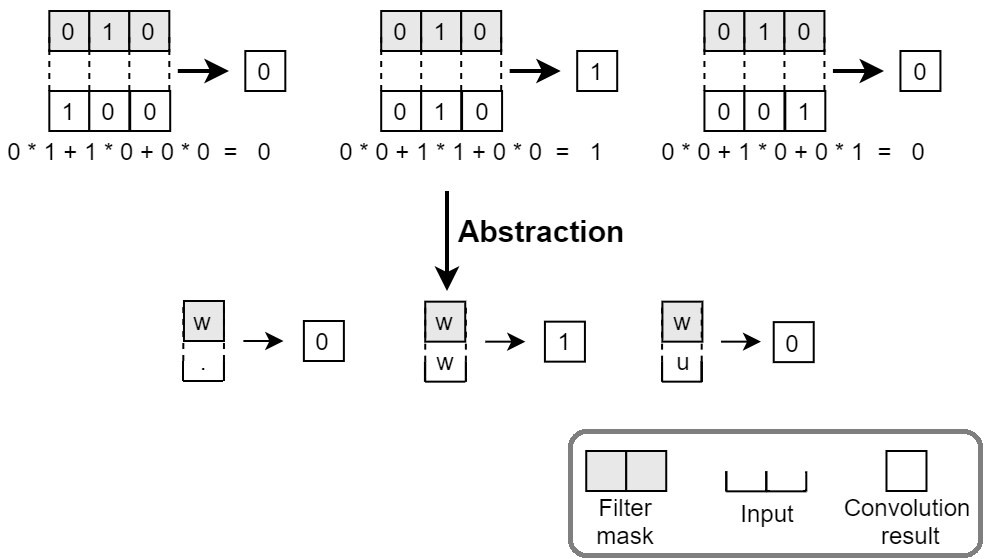}
    \caption{The application of a single CNN filter of size 1 to input}
    \label{fig:cnn filter 1}
\end{figure}
\bigbreak

In the above example, the convolutional filter learns to detect the character 'w', resulting in a 1 output when it hovers above it and a 0 output when it hovers above any other character. Although the example is given for a convolutional filter of size 1, we can easily extend this notation for any filter size, as shown in the upcoming subsections.

\newpage

\subsection{CNN - 'wu' vs 'uw'}
\label{cnn 1}
\textit{Question: should our CNN architecture, or CNN model types in general, be able to distinguish the 'wu' pattern from the 'uw' pattern?} \\

Below we give a simple example of how a CNN model can train a specific mask filter to distinguish between the two patterns successfully. We also note that the use of a max pooling layer is beneficial if not necessary for the successful separation. 

\begin{figure}[!htbp]
    \centering
    \includegraphics[width=15.1cm]{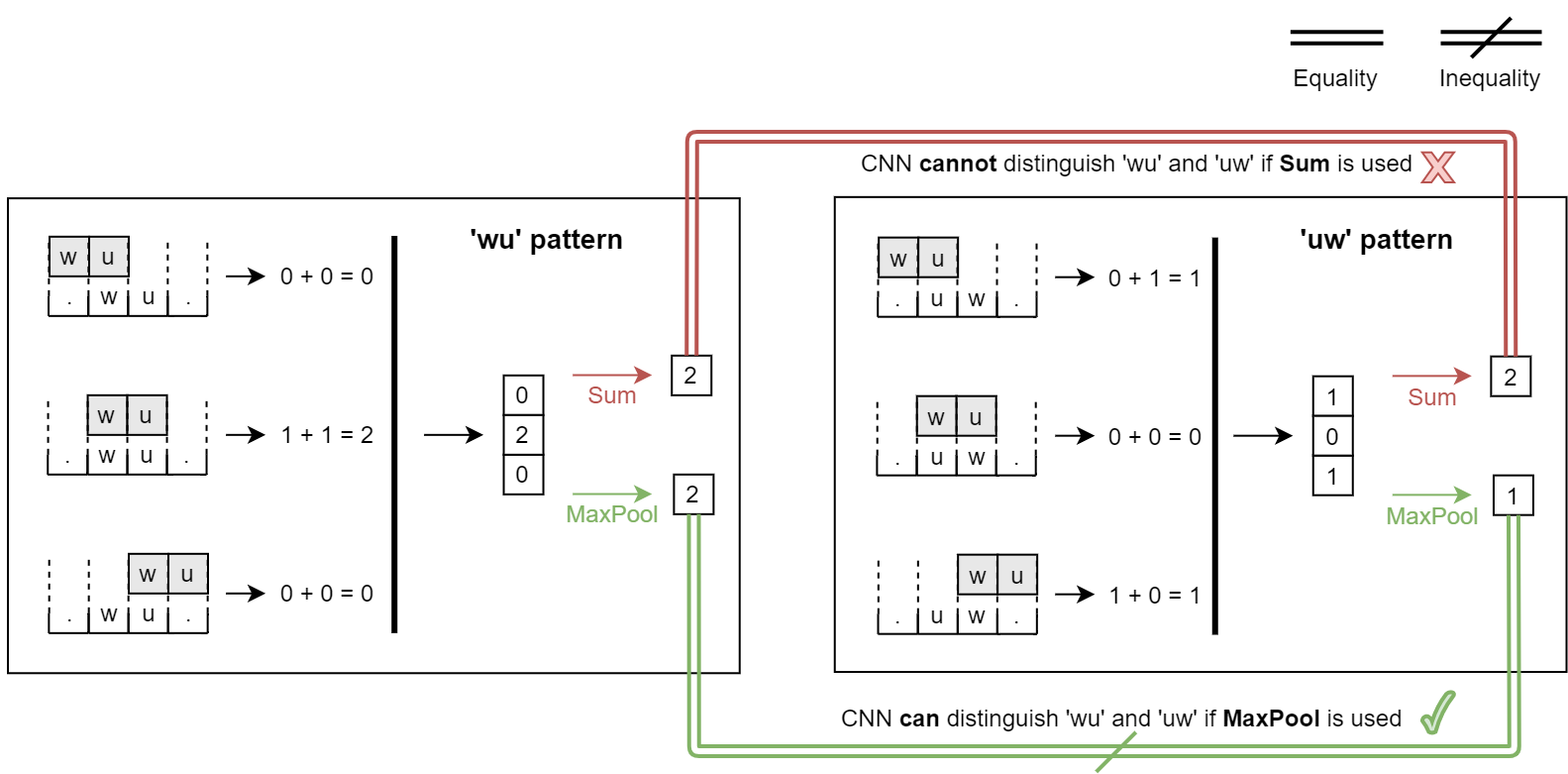}
    \caption{CNN requires MaxPool to distinguish 'wu' from 'uw'}
    \label{fig:cnn filter 2}
\end{figure}

For the sake of simplicity and visualisation, the input dimensions are significantly reduced. We also abstract away the use of padding to maintain the same dimension across the input and first convolutional layer. 

This diagram may raise the question of whether a sum over the convolutional result is necessary (shown in red), and why couldn't the result just be connected to a further dense layer. Indeed, in this minimal example, such a model could successfully learn this pattern without the need for a max pooling layer. However, for larger inputs, we must maintain model robustness to position variance of operations. This can either be done through summing the convolutional result (sum is a permutation invariant function, see DeepSets \cite{zaheer2017deep}), or through a max pooling layer.

Indeed, max pooling layers have been traditionally used for the extraction of the sharpest features \cite{6883}, which is also what enables our CNN's to distinguish these two patterns successfully. Although all of the results concretely discussed in this report feature only CNN's with multiple max-pooling layers, it is indeed our experience that the addition of max pooling layers did improve the overall accuracy of CNN's.

\newpage

\subsection{CNN - position variance}
\label{cnn 2}
\textit{Question: should our CNN architecture, or CNN model types in general, do well on the 2nd filter?} \\

It should not be hard to see how the innate locality of convolutional filters gives the CNN's robustness to the position variance of code. Indeed, because the convolutional filters only consider local neighbourhoods within the input, the absolute position of an operation is not as important. However, what happens in the extreme case of Filter 2, where all meaningful operations can only appear in the first half of programs in the train set and can only appear in the second half of programs in the test set? Indeed, we show that simple convolutional layers may not be enough and a max pooling layer is again required.

\begin{figure}[!htbp]
    \centering
    \includegraphics[width=15cm]{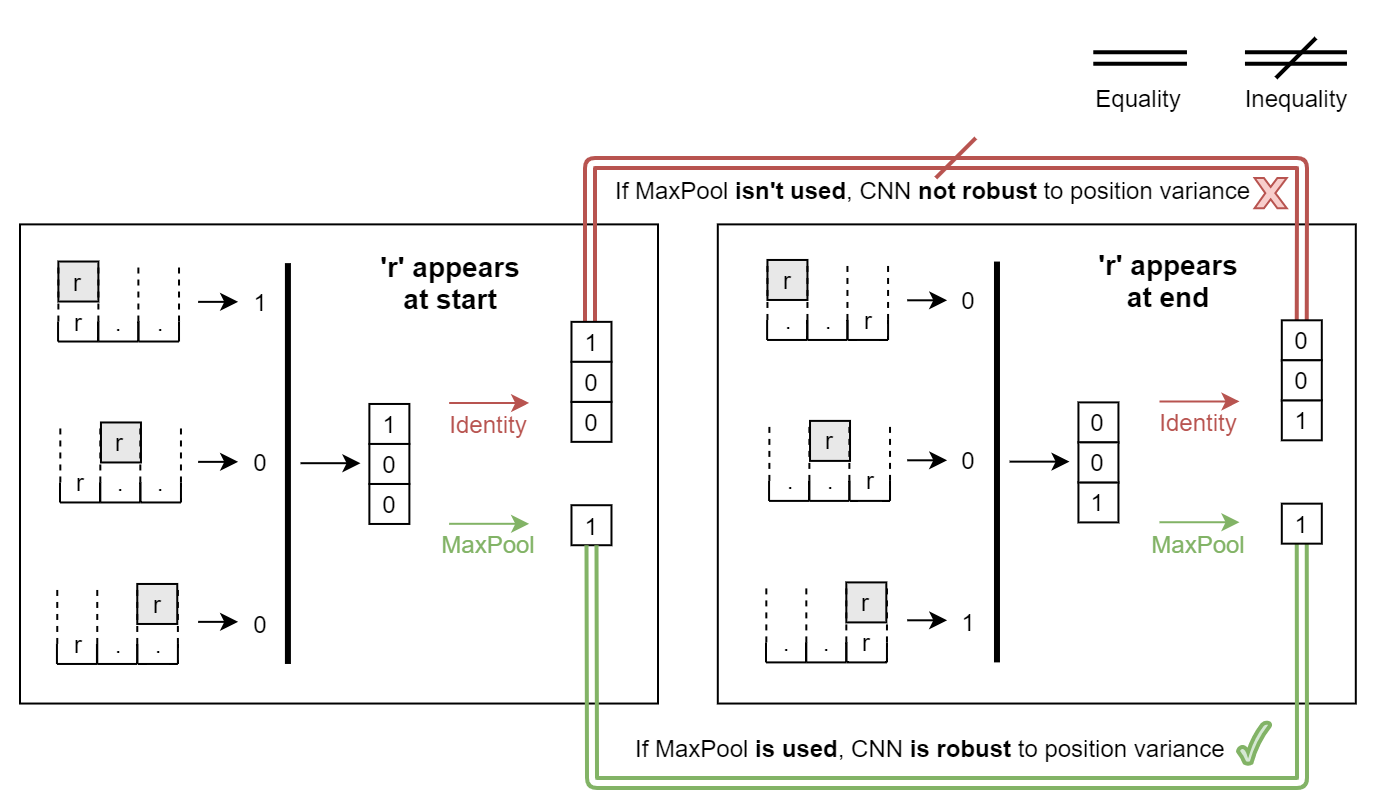}
    \caption{CNN requires MaxPool to do well on Filter 2}
    \label{fig:cnn filter 3}
\end{figure}

As shown above, a max pooling layer unsurprisingly helps with the position variance. It is worth noting that the max pooling needs to be done over the entirety of the first dimension to fully achieve position invariance. In our architecture, this functionality is implemented in a manner very similar to VGGNet \cite{simonyan2014very}, with the continuous chaining of multiple convolutional and max pooling layers. 

Although we do not validate this with the results included in this report, it is also our experience that performing a max pool over the entirety of the first dimension drastically improves the performance of CNN's on Filter 2.

\newpage

\subsection{CNN - inter-operational distance variance}
\label{cnn 3}
\textit{Question: should our CNN architecture, or CNN model types in general, do well on the 4th Filter?}\\

In the previous two subsections, we have provided reasonable examples of useful weights that are well within the capabilities of CNN's to learn. However, we expect that the patterns in Filter 4 go beyond the learning capabilities of CNN's.

Supposing the CNN learns the 'wu' filter mask as shown in the first example \ref{cnn 1}, what happens when the same filter mask is applied to 'w..u' which it has never seen before?

\begin{figure}[!htbp]
    \centering
    \includegraphics[width=15cm]{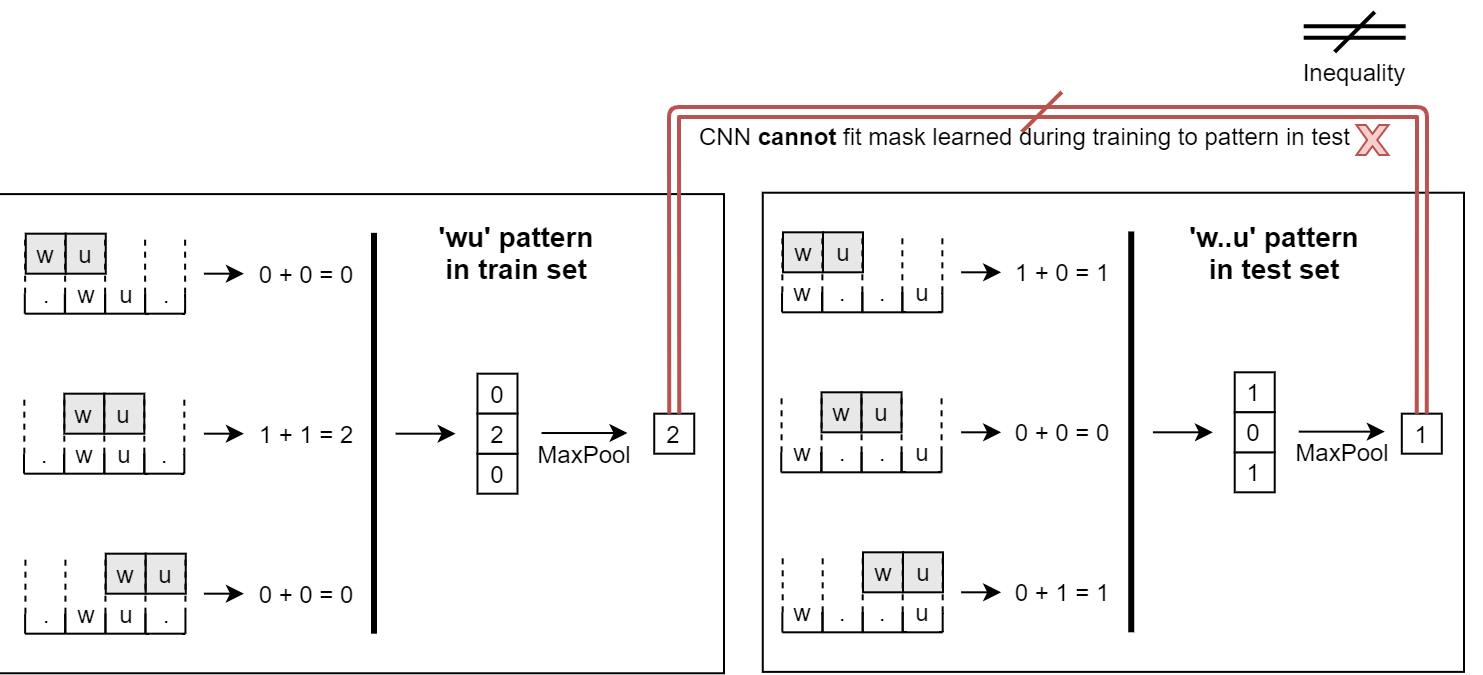}
    \caption{CNN unable to generalise from 'wu' to 'w..u' due to the fixed mask dimensions}
    \label{fig:cnn filter 4}
\end{figure}

\bigbreak

As seen above, the end result for applying the 'wu' filter mask to 'wu' input is different than applying it to 'w..u' input. If we separate all samples containing 'wu' to train and all samples containing 'w..u' to test, we do not expect that a CNN would also train the 'w..u' filter mask. But what if we removed the MaxPool layer and just performed a sum over the convolutional result? Then the model would struggle with distinguishing 'w..u' and 'u..w' \ref{cnn 1} since MaxPool is crucial in that case. \\

Weight regularisation also seems to play a significant role in the case of Filter 4. Empirically, adding $0.01$ L2 regularisation on the 4th Filter experiment helps the CNN's achieve near-perfect accuracy, excepting the 'wu $|$ dr' case. Any patterns other than 'dr' in Function 3 can be learned just by looking at the inclusion of a single operation. Hence CNN's can still achieve relatively good accuracy regardless of the variance in the inter-operational distance.

\newpage

\section{LSTM}
The LSTM model type was the last one added to our roster. We theorised it would be the best performing one, achieving better accuracy than CNN's in the case of the 4th Filter. We show in our results that is indeed to be the case, however, CNN's perhaps unexpectedly perform better on the 2nd Filter.\\

\begin{figure}[!htbp]
    \centering
    \includegraphics[scale=0.32]{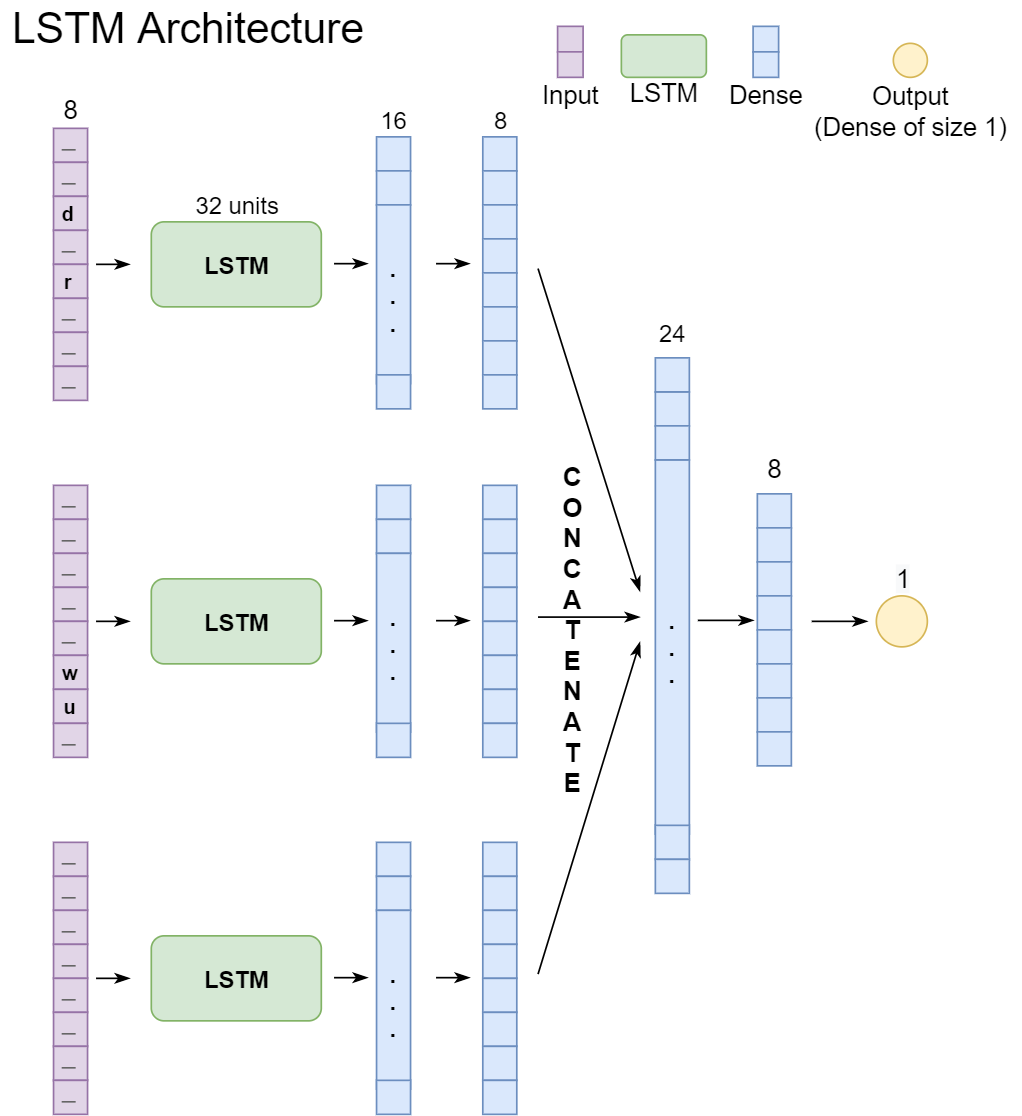}
    \caption{LSTM architecture}
    \label{fig:lstm}
\end{figure}

\bigbreak
\bigbreak

\newpage

\section{Keras vs PyTorch}
The development of the machine learning models initially started with the Keras library \cite{chollet2015keras}, as it provides a very accessible API. However, as models started to gain more architectural complexity (e.g. DeepSets layers), we thought it would be better to permanently move all of our models to the PyTorch library \cite{paszke2017automatic} instead, as it provided more fine-grained control over our model's architectures. To ensure the move from Keras to PyTorch would not impact our models' accuracies, we performed side-by-side comparisons on the accuracy evolution of models throughout epochs on the same dataset. We found that the default Keras weights initialisation ("glorot\_uniform") leads to faster and better convergence than the default PyTorch weight initialisation. Fortunately, we were able to change the PyTorch's weight initialisation to match the Keras one, leading to the models having near identical performance (except for randomness in weight initialisation).

\section{Training}
The character arrays that we generate are one-hot encoded before being fed into the models. At the moment, we have only experimented with the same fixed length for all strings. Still, we do also provide padding functionality for our dataset to deal with variable size input. \\

We train our models for a large number of epochs (typically 100-200), storing train accuracy and loss as well as validation accuracy and loss at each epoch. We utilise two checkpoints, the first one saving the model weights corresponding to the best epoch performance measured as validation accuracy. The second checkpoint performs early stopping if the validation accuracy has not improved after a large number of epochs. Our choice of an optimiser is Adam \cite{goodfellow2016deep}, as empirically it provided the best performance. We also utilise class weights to balance the positive to the negative class ratio (since there are 7 buggy cases and 9 non-buggy cases).

\newpage

\chapter{Evaluation}
The core of the evaluation is represented by how we generate the data and how we separate certain patterns from the training and testing sets. This allows us to test our hypotheses about the capabilities and fundamental limits of each model type.

\section{Formalizing model properties}
In order to thoroughly test the learning capabilities of the models, we need to formally define which properties we want to check, as well as the relevant experiments that need to be carried out. These properties are inspired by patterns one might encounter when modelling real-life programs. We explain why we anticipate some models to perform better than others. We have already discussed the theoretical capabilities of the CNN model type in its own section \ref{cnn}. We provide brief reasoning for the other 3 model types below, as these do not require an in-depth analysis to argue about their behaviour. 

\subsection{Robustness to position variance of code}
If during the training stage a model learns that samples [r...], [.r..] are classified as positive, can it generalise such that during the testing stage [..r.], [...r] are also classified as positive?

    \begin{itemize}
        \item In real programs, an instruction may appear at any position in the program, but it may be that the absolute position at which it appears does not matter. Additionally, we might only see an instruction appear in a limited set of positions during training, but nonetheless want to generalise during testing for the other positions as well.
        
        \item We measure a model's robustness to position variance by taking overall accuracy in the 2nd Filter experiment, as it is the Filter that separates samples in train and test based on operation position
        
        \item Due to their innate locality, we expect that only LSTM's and CNN's should be robust to this. We've already argued CNN's robustness before \ref{cnn 2}. LSTM's should be robust as we think it is both reasonable and within an LSTM's layer ability to learn certain hidden state weights such that noisy operations (',.\_ ') would never affect the current hidden state, leaving only meaningful operations capable of changing the hidden state.
        
        Models with dense layers connected to the input should not be robust to this. This is because the weights corresponding to input positions that are \textbf{not} occupied by meaningful operations during train are not updated (i.e. in our example this corresponds to [..r.] with 'r' at position 2, [...r] with 'r' at position 3) 
    \end{itemize}

\subsection{Robustness to inter-operational distance variance}
 If a model learns that samples [wu], [w..u] are classified as negative, can it generalise such that [w.u] is also classified as negative?
 
    \begin{itemize}
        \item We measure a model's robustness to position variance by taking the accuracy of the 'wu-dr' case for the 4th Filter experiment (all cases other than 'wu-dr' can be learned without having this robustness, as only one operation in Function 3 is needed to make a decision)
        \item Just like in real life, the distance between two operations that jointly affect the shared state of a program can vary, and therefore only some distances might appear during training, but nonetheless we want to generalise for any reasonable distance 
        \item We do not expect Feedforward and DeepSets to be robust to this, as very specific weights are learned during the training stage. The weights that are related to joint operations are thus uniquely determined by both the position of the joint operations in the program as well as the distance between the two operations. Therefore operations with varying distances during the testing stage will not fit the particular weights learned during training.
        
        On the other hand, we expect LSTM's to robust to this, for the same reasons as discussed for the previous property. However, we expect CNN's to fail, as argued before \ref{cnn 3}

    \end{itemize}

\subsection{Ability to distinguish relative ordering of operations}
Can a model distinguish between classifying [w.u] as negative but [u.w] as positive?

    \begin{itemize}
    
        \item We measure a model’s ability to distinguish relative ordering through the following formula:
        
        $score = min(1 - r, r) / 0.5$ 
        
        \quad where $r = dr\_wu / (dr\_wu + dr\_uw)$
        
        \qquad where $dr\_wu$ is the model accuracy for 'dr-wu' case in Filters 1 and 3
        
        \qquad where $dr\_uw$ is the model accuracy for 'dr-uw' case in Filters 1 and 3
        
        \text N.B. both $r$ and $score$ are in the $[0, 1]$ range

        \item If a model correctly classifies 'dr-uw' as buggy but incorrectly classifies 'dr-wu' as also buggy, that means it cannot distinguish between the [u.w] and [w.u] and thus $r$ is minimised (remember 'dr-uw' and 'dr-wu' have different truth labels). Similar reasoning is applied for the vice-versa case, where $1 - r$ is minimised instead
        
        \item On the other hand, $min(1 - r, r)$ is maximised when both accuracies for $dr\_wu$ and $dr\_uw$ are the same so $1 - r = r = 0.5$ (hence the normalizing through dividing by 0.5 to keep the score in the $[0, 1]$ range). Thus the model correctly distinguishes 'dr-uw' as buggy and 'dr-wu' not buggy
        
        \item Filters 1 and 3 are used to check for this property as they do not test for any of the other two properties, therefore allowing us to clearly separate property checks by Filters
        
        \item Another simple, real-life example that reflects this sort of property is represented in the case of assigning to a variable \textbf{after} initialising it (valid \cmark) as opposed to assigning to a variable \textbf{before} initialising it (erroneous \xmark)
        
        \item In theory, all models except DeepSets variation \#2 should be able to learn this since the position of an operation in the input vector uniquely determines the weight associated with it (therefore giving the ability to distinguish two different operation orders). In practice, however, we do not expect that it is feasible for Feedforward and DeepSets variation \#1 to learn this due to a large number of possible combinations (have to distinguish [wu] and [uw] when they occur at position 0 in the program, when they occur at position 1 and so forth). 
        
        However, because of the innate locality of CNN's and LSTM's, the number of combinations is significantly smaller and does not pose an issue, hence we expect them to distinguish the two cases. We've already shown how CNN's can learn to distinguish relative ordering \ref{cnn 1}. It is not hard to see why LSTM's should learn to distinguish this. For example, LSTM's can learn this pattern by maintaining a zeroed hidden state until the first 'u' is met. When a 'w' is met instead, the hidden state is negated. When noisy operations are met, the hidden state remains unchanged. Since negating a zeroed state results in the same zeroed state, 'wu' results in a non-zeroed output state and 'uw' results in a negated non-zeroed output state, hence the difference between the two cases

    \end{itemize}

\section{Testing procedure}
We take several steps to ensure that when testing for model capabilities, we are not affected by randomness. We train each model for a large number of epochs (100-200) as some models might get unfavourable weight initialisations. For each model type, we also train several instances of that type (20 models per type) so as to ensure statistically meaningful results (the only difference between these model instances are the initial weights). Finally, we look at both the aggregates for all the models and also the aggregates for the top 50\% best performing models (aggregates are in either case grouped by model types). 

\section{Iterative experiments}

As mentioned before, we improve our understanding of the problem by iterating on previous experiments as we fine-tune our models and increase or decrease the complexity of our synthetic dataset and its Filters. Within the first series of experiments, we were only testing the Feedforward and Deepsets model types and the process was overall less planned. As we started to gain a better understanding of the problem, we would add other model types and approach the experimentation process in an orderly manner. We could start anticipating results, like how CNN's and LSTM's would outperform their counterparts on Filter 1.

Towards the later stages of the project, we started formally defining the properties we wanted to check for and how to assess them. Within the first round of formal experimentation, we obtained anticipated results for the most part, excepting LSTM performance on Filter 2 and CNN performance on Filter 4. We remedied the LSTM's performance on Filter 2 by adding noise padding to the left of the input during the training stage, and noise padding to the right during testing. This ensured that both during train and test the meaningful operations would be situated in the middle of the program. On the other hand, we increased the weight regularisation for the CNN's in Filter 4 and reduced the number of filters per convolutional layer, leading to a much better performance than Feedforward and DeepSets.

Improving LSTM and CNN performance led us to conduct the last series of experiments. For each Filter, we conducted two separate versions of experiments of varying difficulties. Because each Filter comes with its own parameters that control how many samples appear during training vs. testing (e.g. train-to-test ratio in Filter 1, skip size in Filter 3), we can calibrate the difficulty of the experiments. Indeed, although both the easy and hard version are consistent in results with each other, the differences between model types become more evident in the easy version of the experiments. This is because the easy version is more focused on testing the limits of the models as opposed to the likelihood of the models achieving those limits during training. Therefore, the following results are based on the easier Filters versions. However, we include in the appendix all results.

\newpage
\section{Results}

\subsection{Checking model properties}

As described above, we calculate the scores of each model type for each one of the three properties. Below we provide a table showcasing these results, confirming our initial hypotheses.

\begin{table}[!htbp]
    \centering
    \includegraphics[width=15cm]{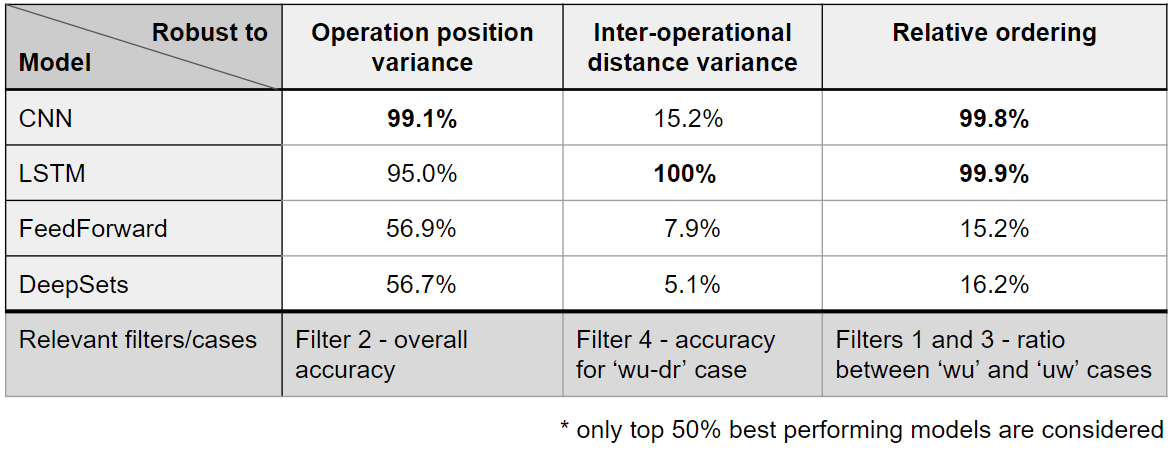}
    \caption{Scores broken down by model types and properties - the larger the score, the better for the model performance}
    \label{fig:pipeline}
\end{table}

As we can see, it is, in fact, the case that mostly due to the innate locality of CNN's and LSTM's they outperform
Feedforward and Deepsets on all of these properties, the sole exception being the poor performance of CNN's on the 4th Filter. However, we had already reasoned this would be the case before running the corresponding experiment \ref{fig:cnn filter 4}. \\

However, one result we did not anticipate was the significant difference in scores between CNN's and LSTM's on Filter 2. We attribute the worse performance of LSTM's to a problem shared with all RNN's, namely the vanishing gradient problem when dealing with longer inputs \cite{hochreiter1997long}. This flaw of RNN's is particularly exacerbated in the more extreme case of Filter 2, where the distributions of the train and test data are vastly different. \\

It is worth noting that Feedforward and DeepSets achieve 56\% on the first property due to the measurement of the score as the overall model accuracy on Filter 2. Therefore, in this case, a score of 56\% is not much better than a score of 50\% corresponding to making random predictions.  \\

We remind that these results are taken from the top 50\% best performing models, on the easy variations of the experiments. Please refer to the appendix for the full results.

\newpage

\subsection{Results by Filters and patterns}

In order to either support some of our earlier claims or provide further arguments, we also include model accuracy broken down by either Filters or patterns.

\begin{table}[!htbp]
    \centering
    \includegraphics[width=15cm]{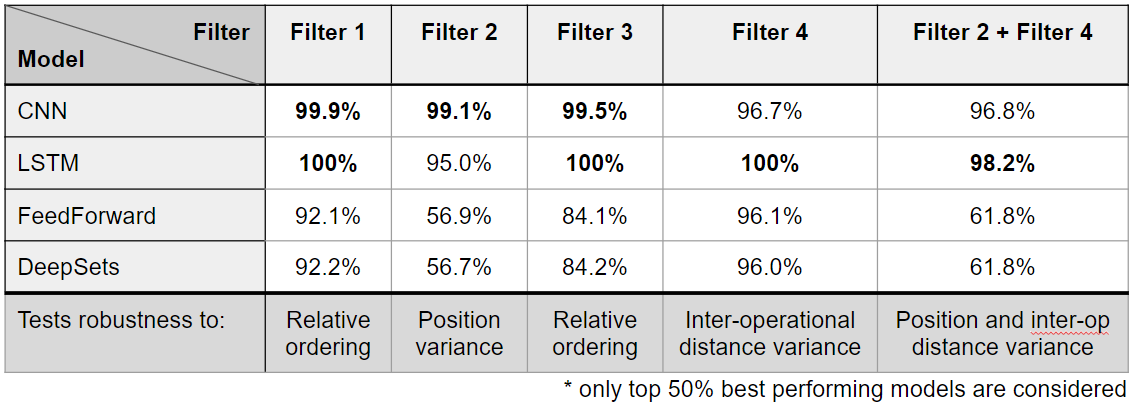}
    \caption{Scores broken down by model types and Filters}
    \label{combiningfiltersresults}
\end{table}

One of our earlier claims was that combining Filters 2 and 4 makes sense empirically \ref{combiningfilters}. Indeed, as we can see for all models the accuracy under the experiment of Filter 2 + 4 is always a weighted average of the accuracy under Filter 2 and the accuracy under Filter 4. This validates the idea the models would never perform worse on the combined Filters than the hardest Filter nor would they perform better on the combined Filters than the easiest Filter. \\

\bigbreak

\begin{table}[!htbp]
    \centering
    \includegraphics[width=15cm]{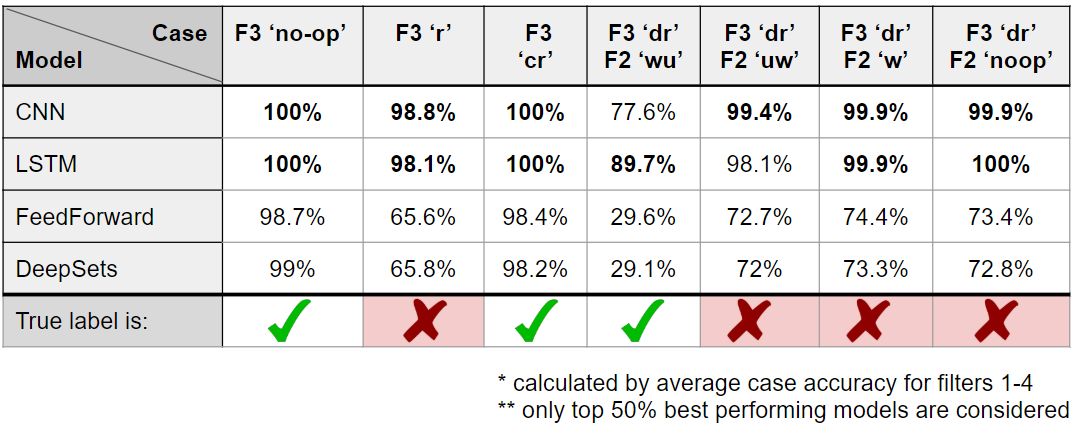}
    \caption{Scores broken down by model types and patterns}
    \label{patternresults}
\end{table}

Moreover, we mentioned that the 'wu $|$ dr' pattern would be the hardest for models to classify successfully \ref{fig:complex rule}. Indeed, as shown above, that seems to be the case with the 'wu $|$ dr' pattern leading to a worse average accuracy than any other pattern, for any of the model types. \\

\newpage

\subsection{Additional comments}

In the Filters section, we discuss how certain parameters can lead to Filter 3 also testing for robustness to position variance \ref{section:filter3}. Indeed, for models that are not robust to position variance (i.e. Feedforward and DeepSets), there is a difference in performance. They achieve better performance when there is no position variance between train and test distributions \ref{filter3} and worse performance when there is some variance \ref{filter3variance}. \\

\begin{figure}[!htbp]
    \centering
    \includegraphics[width=15cm]{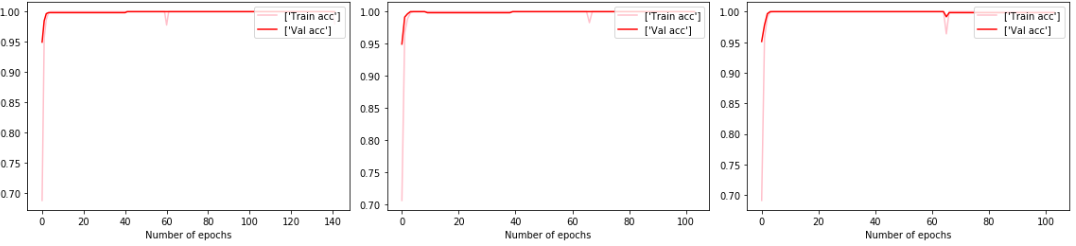}
    \caption{Convergence of CNN's under Filter 1}
    \label{cnnplot}
\end{figure}

\begin{figure}[!htbp]
    \centering
    \includegraphics[width=15cm]{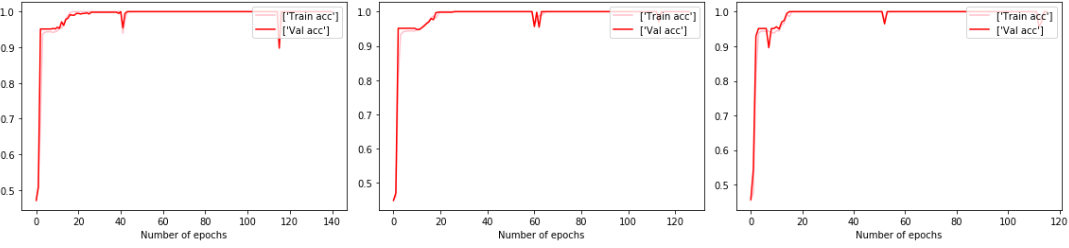}
    \caption{Convergences of LSTM's under Filter 1}
    \label{lstmplot}
\end{figure}

Additionally, we monitor the time it takes to train each model type as well as the performance at each epoch. There exist differences between the training times of model types (for e.g., on a GPU in the 2nd Filter experiment CNN's and LSTM's take 1h49min and 1h42min while Feedforward and DeepSets take 27min and 26min respectively) as well as epochs till convergence (for e.g., CNN's \ref{cnnplot} have a slight edge over LSTM's \ref{lstmplot} under Filter 1). However, in our experience, we do not find that the discrepancies between the good models are significant enough to justify using one model over the other. \\

\newpage

\chapter{Conclusion}
\section{Summary}
In conclusion, we provide a synthetic dataset meant to simulate patterns found in real-life (concurrent) programs. We also provide filters for this dataset, meant to probe the models on certain inputs so as to test certain properties. We reason about the fundamental limits and capabilities for each model type, spending time in particular to explore CNN mechanisms. We formalise certain desirable properties we might want to have in neural networks when applied to real programs. We then empirically show that our initial expectations regarding the learning limits or capabilities of models are met, presenting the advantages and disadvantages of various model types on this problem. We show that models that have a local view of the input (CNN's, LSTM's) have an advantage over those that do not (Feedforward, DeepSets). We also show models traditionally used for other sequential problems (LSTM's) may also outperform close contenders (CNN's). We hope our findings provide more insight into the performance of neural networks on sequential problems as a whole.

\section{Future work}
In the specific context of comparing various model types and learning to model code as input features (i.e. the contributions of this SuperUROP project), there are several immediate issues that we would like to tackle next. 

\begin{itemize}

    \item \textbf{Exploration of graph neural networks (GNN's)}
    
    When modelling real code, we will most likely use a graph representation as input as opposed to the vector approach that we take in our synthetic dataset. This is due to the high variance in program size and number of concurrent sections, which makes the modelling of input as done in this report perhaps unfeasible on real programs. Much like RNN's, GNN's perform well on variable sized input. Additionally, they can utilise gating mechanisms specific to the RNN model types, which we've shown to work well on our synthetic dataset.

    \item \textbf{Further model interpretation}
    
    Within this project we've used the separation of certain sample sets into train and test as a means of interpreting the learning limits and capabilities of a model. In the future, we would like to have a more detailed look at our models, particularly in terms of input feature importance \ref{section:2.3} and layer weights \ref{layerweights}.
    
    \item \textbf{Adding further complexity to the dataset and filters}
    
    As a means of closing the gap between our synthetic problem and the real use case, we need to further increase the complexity of our simulated programs. This could be done by increasing the complexity of our grammar through adding more operations to our Functions (perhaps in particular to Function 1). Additionally, we could also add an additional Filter, which would separate based on the number of operations in each Function. While for most model types we would have to pad the input to normalise the input length, LSTM's would not require any further input preprocessing, hence offering another potential advantage.

\end{itemize}

\bigbreak
\bigbreak

Within the broader aim of the project (i.e. developing concurrency tools with neural networks), we mention some of the most important milestones that we will need to achieve.

\begin{itemize}
    \item \textbf{Predicting buggy operations with preciseness}
    
    It is not enough for a model to simply classify a program as buggy or not. The programmer also needs to know which specific lines of code are causing the issue. This problem may be non-trivial and have some overlap with the "Further model interpretation" keypoint presented above.
     
    \item \textbf{Figuring out the intermediate representation}
    
    We have to keep in mind one of the main motivations of our project, which is the support of newer languages through training our models on an intermediate representation common to all source languages. One key factor to look out for is the translation from a newer source language to the intermediate representation, which could require additional implementation effort.
    
    \item \textbf{Procurement of a labelled concurrency dataset}
    
    The most challenging milestone is also the primary motivation for our use of a synthetic dataset. Manual labelling of concurrent programs is very laborious and error-prone, especially when considering multiple source languages. We could use current tools to label a dataset: even though the model will never beat the original tool's performance, achieving comparable results means we successfully modelled concurrent programs. Additionally, we may even achieve better performance by combining datasets with different source languages obtained from multiple tools.
    
\end{itemize}

\bibliography{citation}
\bibliographystyle{ieeetr}

\chapter{Appendix}

\section{Full results}

Here we provide the results used to calculate the aggregate accuracies and scores displayed in the results section. We provide both the easy versions (on the left) and the hard versions (on the right) of all experiments. We remind that the aggregated results from the evaluation chapter are calculated using the EASY versions of the experiments. We also only record the results for the best 50\% performing models.
\bigbreak

\subsection{Filter \#1}

\hspace{-1.9cm}
\begin{minipage}{0.6\textwidth}
\begin{flushleft} \large
\includegraphics[width=\linewidth]{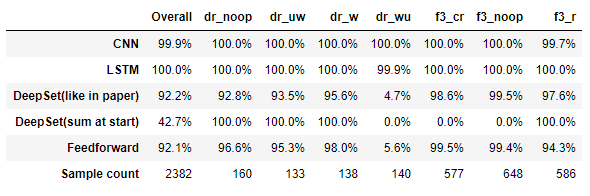}\\
\captionof{table}{Filter 1 EASY version - train subset is 25\% of full dataset}
\end{flushleft}
\end{minipage}
~
\hspace{0.3cm}
\begin{minipage}{0.6\textwidth}
\begin{flushleft} \large

\includegraphics[width=\linewidth]{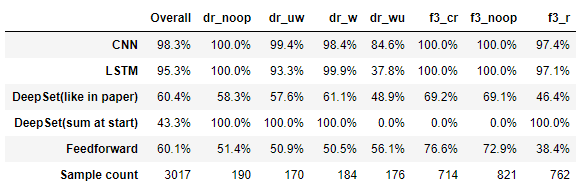}\\
\captionof{table}{Filter 1 HARD version - train subset is 5\% of full dataset}
\end{flushleft}
\end{minipage}

\bigbreak
\bigbreak
\bigbreak

\subsection{Filter \#2}

\hspace{-1.9cm}
\begin{minipage}{0.6\textwidth}
\begin{flushleft} \large
\includegraphics[width=\linewidth]{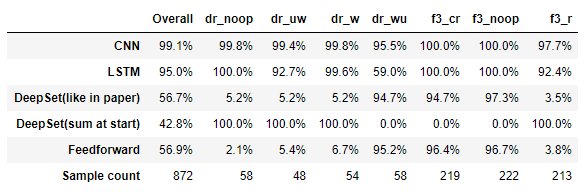}\\
\captionof{table}{Filter 2 EASY version - train set has operations appearing the first half of the program, test set in second half}
\end{flushleft}
\end{minipage}
~
\hspace{0.3cm}
\begin{minipage}{0.6\textwidth}
\begin{flushleft} \large

\includegraphics[width=\linewidth]{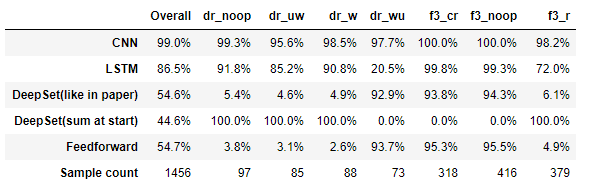}\\
\captionof{table}{Filter 2 HARD version - train set has operations appearing the first third of the program, test set in last two thirds}
\end{flushleft}
\end{minipage}

\bigbreak
\bigbreak
\bigbreak

\subsection{Filter \#3}
\label{filter3}

\hspace{-1.9cm}
\begin{minipage}{0.6\textwidth}
\begin{flushleft} \large
\includegraphics[width=\linewidth]{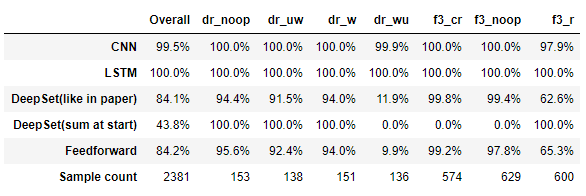}\\
\captionof{table}{Filter 3 EASY version - every 4th sample is allocated to the train set}
\end{flushleft}
\end{minipage}
~
\hspace{0.3cm}
\begin{minipage}{0.6\textwidth}
\begin{flushleft} \large

\includegraphics[width=\linewidth]{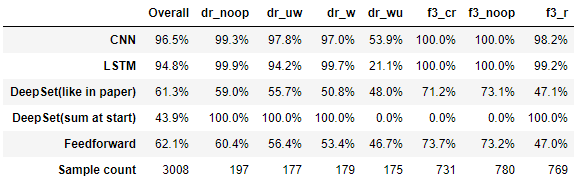}\\
\captionof{table}{Filter 3 HARD version - every 19th sample is allocated to the train set}
\end{flushleft}
\end{minipage}

\bigbreak
\bigbreak

\subsection{Filter \#3 - position variance}
\label{filter3variance}

\hspace{-1.9cm}
\begin{minipage}{0.6\textwidth}
\begin{flushleft} \large
\includegraphics[width=\linewidth]{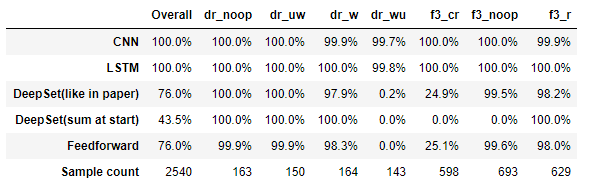}\\
\captionof{table}{Filter 3 (with position variance) EASY version - every 5th sample is allocated to the train set}
\end{flushleft}
\end{minipage}
~
\hspace{0.3cm}
\begin{minipage}{0.6\textwidth}
\begin{flushleft} \large

\includegraphics[width=\linewidth]{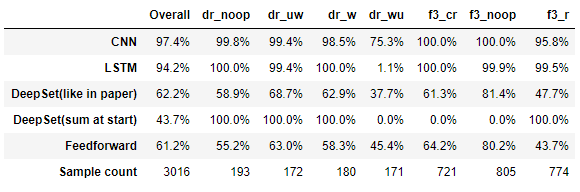}\\
\captionof{table}{Filter 3 (with position variance) HARD version - every 20th sample is allocated to the train set}
\end{flushleft}
\end{minipage}

\bigbreak
\bigbreak

\subsection{Filter \#4}

\hspace{-1.9cm}
\begin{minipage}{0.6\textwidth}
\begin{flushleft} \large
\includegraphics[width=\linewidth]{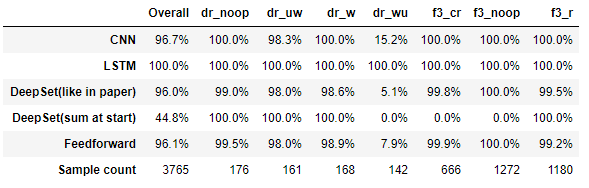}\\
\captionof{table}{Filter 4 EASY version - distances between meaningful operations are 0 and 2 in train, 1 in test}
\end{flushleft}
\end{minipage}
~
\hspace{0.3cm}
\begin{minipage}{0.6\textwidth}
\begin{flushleft} \large

\includegraphics[width=\linewidth]{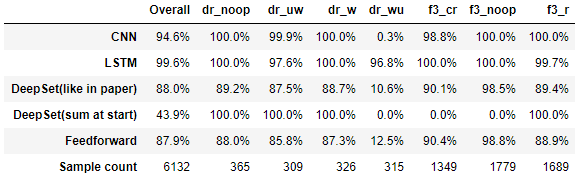}\\
\captionof{table}{Filter 4 HARD version - distances between meaningful operations are 1 in train, 0 and 2 in test}
\end{flushleft}
\end{minipage}

\bigbreak
\bigbreak

\subsection{Filter \#2 + \#4}

\hspace{-1.9cm}
\begin{minipage}{0.6\textwidth}
\begin{flushleft} \large
\includegraphics[width=\linewidth]{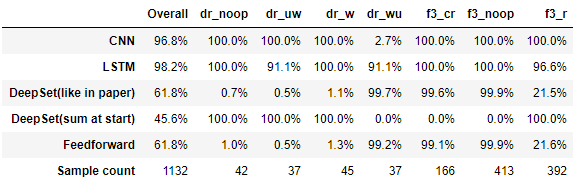}\\
\captionof{table}{Filters 2 and 4 EASY version - combining the EASY version of Filter 2 with the EASY version of Filter 4}
\end{flushleft}
\end{minipage}
~
\hspace{0.3cm}
\begin{minipage}{0.6\textwidth}
\begin{flushleft} \large

\includegraphics[width=\linewidth]{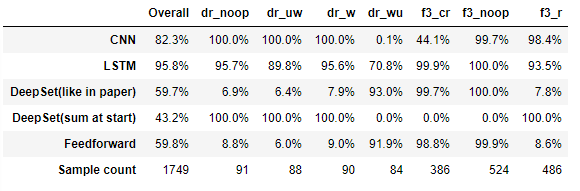}\\
\captionof{table}{Filters 2 and 4 HARD version - combining the EASY version of Filter 2 with the HARD version of Filter 4}
\end{flushleft}
\end{minipage}

\end{document}